\documentclass[nofootinbib,twocolumn]{revtex4-1}

\usepackage{amsmath}
\usepackage{amsfonts}
\usepackage{amssymb}
\usepackage{graphicx}

\usepackage{color}
\usepackage{verbatim}
\usepackage{simplemargins}

\newcommand{\D}{\mathrm{d}}

\newcommand{\tar}{\mathrm{t}}
\newcommand{\pro}{\mathrm{p}}

\newcommand{\idx}{i}

\newcommand{\fix}{\mathrm{fix}}
\newcommand{\cm}{\mathrm{cm}}

\newcommand{\U}{\mathrm{U}}
\newcommand{\T}{\Theta}

\newcommand{\X}{\chi}
\newcommand{\R}{\varrho}

\newcommand{\atan}{\mathrm{arctan}}
\newcommand{\ash}{\mathrm{arsinh}}

\newcommand{\lm}{\mathrm{lim}}

\newcommand{\pull}{\vspace*{-5mm}}
\newcommand{\pullc}{\vspace*{-1mm}}

\setbottommargin{2.2cm}

\begin{document}

\thispagestyle{empty}

\onecolumngrid
\begin{center}

\textbf{\large A shadow of the repulsive Rutherford scattering in the fixed-target and the center-of-mass frame}\\[.5cm]

Petar \v{Z}ugec$^{1*}$ and Ivan Topi\'{c}$^2$\\[.1cm]
{\small
{\itshape
$^1$Department of Physics, Faculty of Science, University of Zagreb, Zagreb, Croatia\\
$^2$Archdiocesan Classical Gymnasium, Zagreb, Croatia\\}
$^*$Electronic address: pzugec@phy.hr\\[.5cm]
%(Dated: \today)\\[.5cm]
}

%{438pt}
\begin{minipage}{400pt}
\small
The paper explores the shadow of the repulsive Rutherford scattering---the portion of space entirely shielded from admitting any particle trajectory. The geometric properties of the projectile shadow are analyzed in detail in the fixed-target frame as well as in the center-of-mass frame, where both the charged projectile and the charged target cast their own respective shadows. In both frames the shadow is found to take an extremely simple, paraboloidal shape. In  the fixed-target frame the target is precisely at the focus of this paraboloidal shape, while the focal points of the projectile and target shadows in the center-of-mass frame coincide. In the fixed-target frame the shadow takes on a universal form, independent of the underlying physical parameters, when expressed in properly scaled coordinates, thus revealing a natural length scale to the Rutherford scattering.\\
\end{minipage}

\end{center}

\twocolumngrid

\renewcommand{\theequation}{\arabic{equation}}

\section{Introduction}
\label{intro}

Rutherford scattering---a scattering of electric charges due to the Coulomb interaction, whether it be attractive or repulsive---is one of the most famous concepts in all of physics. The series of historical experiments by Geiger and Marsden \cite{geiger1,geiger2,geiger3}, demonstrating some as yet unexpected properties in the scattering of $\alpha$-particles by thin metal foils, has lead to a discovery of the atomic nucleus by Rutherford \cite{rutherford} and a subsequent birth of nuclear physics. Today Rutherford scattering is a regular subject of all textbooks on classical mechanics and the introductory nuclear physics, and a basis of several experimental techniques such as Rutherford Backscattering Spectrometry (RBS) \cite{rbs} and Elastic Recoil Detection Analysis (ERDA) \cite{erda}.

\begin{figure}[b!]
\centering
%\vspace*{-3mm}
\includegraphics[width=0.5\textwidth,keepaspectratio]{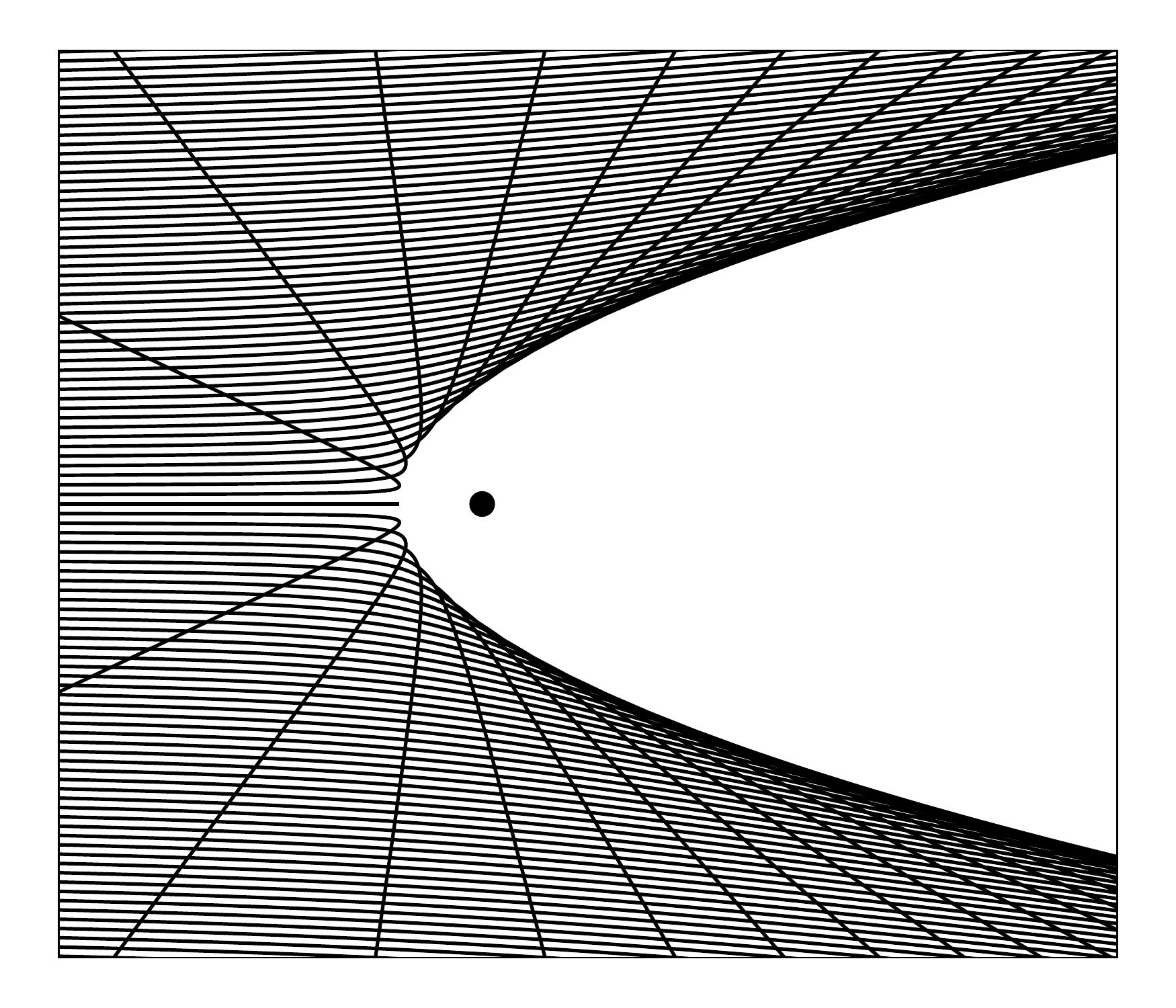}
\pull
\caption{Isoenergetic projectile trajectories in a repulsive Coulomb field of the charged target (the dot), in the fixed-target frame.}
\label{fig1}
\end{figure}

Throughout most of the educational sources, undergraduate or otherwise, the fact that the \textit{repulsive Rutherford scattering casts a shadow} seems to be little known---as if it were entirely forgotten, neglected or ignored. At best, it is only tacitly recognized, whenever a plot such as the one from figure~\ref{fig1} is presented, showing several stacked trajectories for a charged projectile moving through a Coulomb field of a stationary target. Clearly, there is an isolated portion of space admitting no trajectories due to their deflection in a Coulomb field, which can be considered as a proverbial shadow of the repulsive scattering. The attractive scattering of opposing charges shows no such feature, as the trajectories of a projectile of any initial energy may be continuously brought closer to a target, until they bend entirely around it, sweeping out the entire geometric space.

The reality of this shadow is not only of a of great educational value, as a motivation for some beautiful calculations and insights, but also plays an important role in the Low-Energy Ion Scattering Spectroscopy (LEIS) \cite{leis1,leis2}. It may already be intuited from figure~\ref{fig1} that the shape of the shadow in the fixed-target frame might be parabolic, that is paraboloidal when considered in a three-dimensional space. Indeed, the paraboloidal shape is obtained even in a small-angle scattering (i.e. high projectile energy) approximation \cite{shadow0}, leading to the correct value of the paraboloid stiffness parameter (its leading coefficient). This result is often quoted, and its use is justified by the fact that for the low-energy ions not much would be gained by the exact result, as the screening of the `naked' nuclear potential plays a far greater role than the nature of approximation \cite{screen}.

However, rarely is the form of this shadow treated accurately, or analyzed in detail as a subject in its own right, even in a fixed-target frame, let alone in any other reference frame. One example is a mechanics textbook by Sommerfeld, where the topic appears as a supplemental problem I.12, with a short guideline on how to obtain the correct solution \cite{shadow1}. Further examples include succinct works by Adolph \textit{et al.} \cite{shadow_prb1} and Warner and Huttar \cite{shadow_prb2}, all these efforts being decades apart. In the review paper by Burgd\"{o}rfer the paraboloidal shadow form is also quickly obtained, but relying on the properties of the Coulomb continuum wavefunctions \cite{shadow2}. Interestingly, Samengo and Barrachina \cite{shadow_hyp} consider the incident projectile trajectories from a \textit{point source}, finding in that case the hyperboloidal shadow. At the same time, they recover the more familiar paraboloidal shape in the limit $R\to\infty$, with $R$ as the initial projectile-target distance. Most of these works make a point of this easily accessible topic being forgotten \cite{shadow_prb1}, generally unknown \cite{shadow_prb2} and commonly omitted from the standard textbooks \cite{shadow_hyp}. Nothing much seems to have changed to this day.

%The other is the review paper by Burgd\"{o}rfer, where the shadow form is quickly obtained, but relying on the properties of the Coulomb continuum wavefunctions \cite{shadow2}.

We aim to give the subject our full attention and a detailed treatment it deserves, with an intent of rekindling the general interest in it as a worthy educational topic. In this work we analyze the geometric shape of the Rutherford scattering shadow, as it appears in the fixed-target and the center-of-mass frame. A transition to the laboratory frame (where the target is at rest only at the initial moment, i.e. when the projectile is just put into motion) or any other comoving frame (moving with a constant velocity relative to the center-of-mass frame) is much more involved than may seem at the first glance and will be the subject of a separate work. %\cite{lab}

%For completeness, readers' convenience and in order to establish the concepts, constants and conventions used throughout this work, we will give the full derivation of the classical trajectories within the repulsive Coulomb field.

We will treat the scattering kinematics nonrelativistically, which is a common enough approach. There is a myriad of sources offering some form of derivation of the classical trajectories within the Coulomb field, of which we cite only a few classics \cite{kittel,goldstein,landau,spiegel}. The trajectories are usually determined within the context of the two-body Kepler problem, i.e. assuming the gravitational potential, while the chapters on Rutherford scattering typically focus on the scattering cross section. Quite often the problem is approached from the onset with an assumption of a large disparity between the masses, such that one body or particle (e.g. star or the target nucleus) is much heavier than the other (e.g. planet or the charged projectile). The obvious advantage of this approach is that the fixed-target frame also corresponds (at least approximately) both to the center-of-mass and the laboratory frame. As we are interested in each of these frames in its own right, we will follow the more general approach, making no specific assumptions about the masses involved.

Since the fixed-target frame of a finite-mass target is accelerated (as the target will recoil from the incoming projectile), a savvy reader might pose a legitimate question: is the force that the target exerts upon the projectile in such frame purely electrostatic? Or is there also a radiative component to the target's electromagnetic field, even in the frame where it stays at rest, due to its acceleration in an outside inertial frame? The question is, in fact, entirely nontrivial and has indeed been a subject of a long-standing debate. The paradox has since been resolved and we know now that the electric field of a charge at rest is indeed purely electrostatic even in an accelerated frame \cite{paradox}. Therefore, we are entirely justified in assuming the pure electrostatic force between the finite-mass particles in a fixed-target frame, which will serve as a starting point for a transition into any other reference frame.\newpage

This paper is accompanied by the Supplementary note, expanding upon the main material presented herein.

%(to be accessed from the Journal website)

\section{Coulomb trajectories}
\label{trajectories}

Let us consider the Coulomb force $\mathbf{F}_{\tar\rightarrow\pro}$ exerted upon the projectile by the charged target:
\begin{equation}
\mathbf{F}_{\tar\rightarrow\pro}=\frac{Z_\pro Z_\tar e^2}{4\pi\epsilon_0}\frac{\mathbf{r}_\pro-\mathbf{r}_\tar}{|\mathbf{r}_\pro-\mathbf{r}_\tar|^3},
\end{equation}
where $Z_\pro$ and $Z_\tar$ are the projectile and target charge, respectively, in units of the elementary charge $e$; $\epsilon_0$ is the vacuum permittivity; $\mathbf{r}_\pro$ and $\mathbf{r}_\tar$ are the particle and target position-vectors. For each particle the Newton's second law, in combination with the third one (\mbox{$\mathbf{F}_{\tar\rightarrow\pro}=-\mathbf{F}_{\pro\rightarrow\tar}$}), states:
\begin{eqnarray}
&m_\pro\ddot{\mathbf{r}}_\pro=\mathbf{F}_{\tar\rightarrow\pro},
\label{rp}\\
&m_\tar\ddot{\mathbf{r}}_\tar=-\mathbf{F}_{\tar\rightarrow\pro},
\label{rt}
\end{eqnarray}
with $m_\pro$ and $m_\tar$ as the projectile and target mass, respectively. By introducing the center-of-mass position $\mathbf{R}$:
\begin{equation}
\mathbf{R}\equiv\frac{m_\pro\mathbf{r}_\pro+m_\tar\mathbf{r}_\tar}{m_\pro+m_\tar}
\label{R}
\end{equation}
and summing~(\ref{rp})~and~(\ref{rt}):
\begin{equation}
m_\pro\ddot{\mathbf{r}}_\pro+m_\tar\ddot{\mathbf{r}}_\tar=(m_\pro+m_\tar)\ddot{\mathbf{R}}=\mathbf{0},
\end{equation}
one immediately obtains the equation of motion for the center of mass: $\ddot{\mathbf{R}}=\mathbf{0}$, clearly showing that in the absence of the additional external forces the system as the whole can not accelerate, which is to say that the linear momentum of the isolated system is conserved. By subtracting the acceleration terms from~(\ref{rp})~and~(\ref{rt}) and defining the target-relative projectile position $\mathbf{r}$:
\begin{equation}
\mathbf{r}\equiv\mathbf{r}_\pro-\mathbf{r}_\tar,
\label{r}
\end{equation}
one obtains the second equation of motion:
\begin{equation}
\ddot{\mathbf{r}}=\left(\frac{1}{m_\pro}+\frac{1}{m_\tar}\right)\mathbf{F}_{\tar\rightarrow\pro}=\frac{Z_\pro Z_\tar e^2}{4\pi\epsilon_0\mu}\frac{\hat{\mathbf{r}}}{r^2},
\label{eom}
\end{equation}
where the common definition of the reduced mass $\mu$:
\begin{equation}
\frac{1}{\mu}\equiv\frac{1}{m_\pro}+\frac{1}{m_\tar}
\end{equation}
has been used, together with the vector norm $r=|\mathbf{r}|$ and the corresponding unit direction $\hat{\mathbf{r}}=\mathbf{r}/r$.

\begin{figure}[t!]
\centering
\includegraphics[width=0.3\textwidth,keepaspectratio]{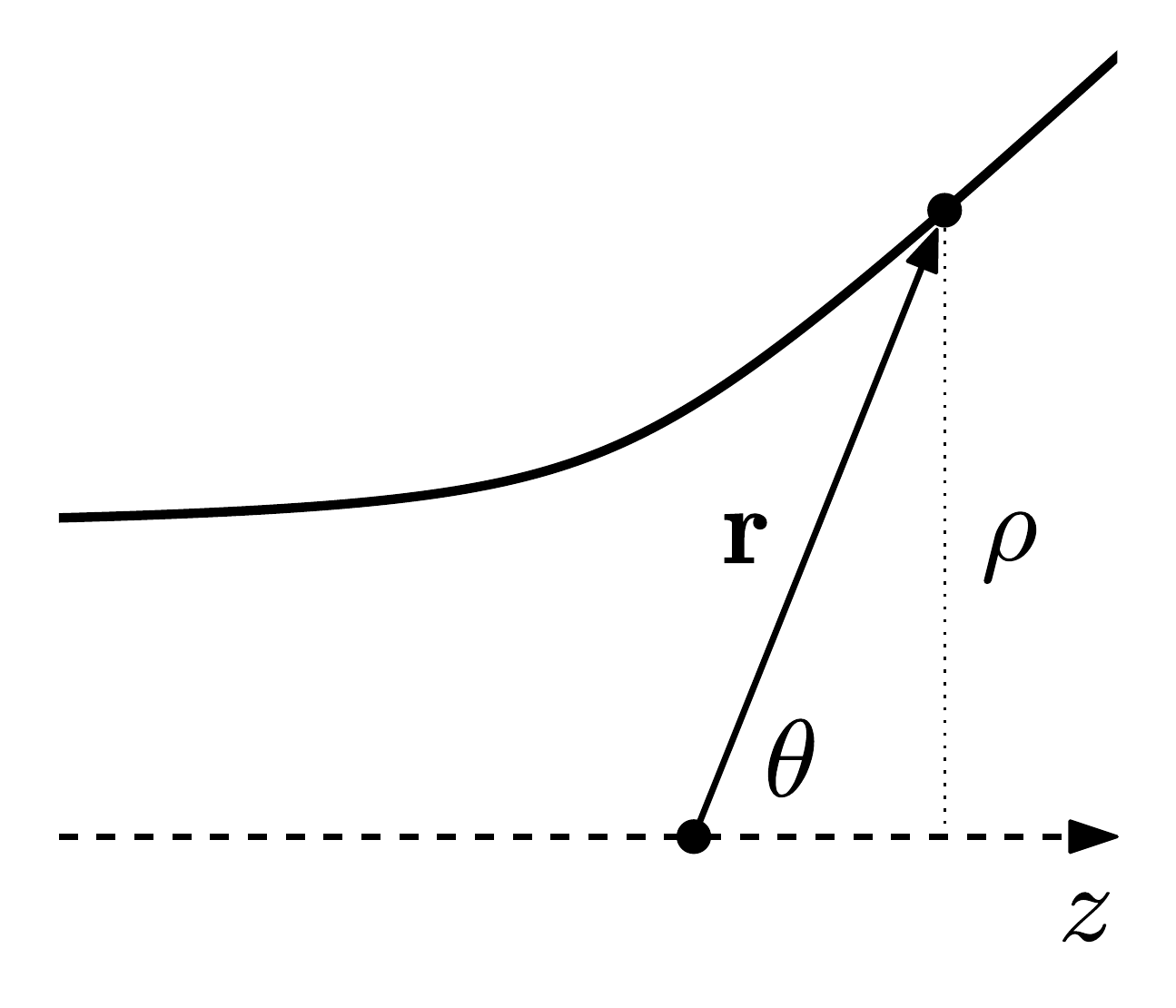}
\pull
\caption{Geometric parameters used for describing the projectile trajectory (full line). The target-relative projectile position $\mathbf{r}$ is described both by the spherical coordinates $r$ and $\theta$, and their cylindrical counterparts $\rho$ and $z$. Due to the axial symmetry the azimuthal angle $\varphi$ bears no relevance to the problem.}
\label{fig2}
\end{figure}

In parameterizing the projectile trajectory we will make use of the cylindrical coordinates $\rho$ and $z$, alongside their spherical counterparts $r$ and $\theta$. Figure~\ref{fig2} clearly illustrates their relation. In that, the direction of the $z$-axis corresponds to the projectile's initial direction of motion (i.e. its initial velocity). Assuming that the projectile has been put into motion as a free particle of initial speed $v_0$ and with the impact parameter  $\R_0$---from the negative side of the $z$-axis, at the infinite distance from the target ($\theta_0=\pi$)---for the initial conditions we can write:
\begin{eqnarray}
&\mathbf{r}(\theta_0=\pi)=\R_0\hat{\boldsymbol{\rho}}+\Big(\lim_{z_0\to-\infty}z_0\Big)\hat{\mathbf{z}},
\label{r0}\\
&\dot{\mathbf{r}}(\theta_0=\pi)=v_0\hat{\mathbf{z}}.
\label{v0}
\end{eqnarray}
Section~A of the Supplementary note derives a solution to the equation of motion~(\ref{eom}) under these conditions. Introducing the following shorthand:
\begin{equation}
\X\equiv\frac{Z_\pro Z_\tar e^2}{4\pi\epsilon_0\mu v_0^2},
\label{x}
\end{equation}
the solution for the relative coordinate may be expressed as:
\begin{equation}
r(\theta;\R_0)=\frac{\R_0^2}{\sqrt{\X^2+\R_0^2}\sin[\theta-\atan(\X/\R_0)]-\X}.
\label{master}
\end{equation}
It is well known that for the repulsive interaction, i.e. for $\X>0$ (\ref{master}) defines a hyperbolic trajectory. By construction, one of its asymptotes is parallel to the $z$-axis, at a distance $\R_0$ from it. The other asymptote is defined by the famous scattering angle $\vartheta$  from a fixed-target frame, which is easily determined from (\ref{master}) as the angle for which the expression diverges, i.e. the denominator vanishes, leading to:
\begin{equation}
\vartheta=2\:\atan\frac{\X}{\R_0}.
\label{scatangle}
\end{equation}
This also means that some particular angle $\theta$ may be reached only by those trajectories whose scattering angle is further on ($\vartheta<\theta$). Evidently, those are the trajectories whose impact parameter satisfies: $\R_0>\X/\tan(\theta/2)$.

It should be noted that (\ref{master}) has a universal shape, independent of all the underlying physical parameters borne by $\X$, when expressed in the scaled, dimensionless coordinates \mbox{$\bar{r}=r/\X$} and \mbox{$\bar{\R}_0=\R_0/\X$}:
\begin{equation}
\bar{r}=\frac{\bar{\R}_0^2}{\sqrt{1+\bar{\R}_0^2}\sin[\theta-\atan(1/\bar{\R}_0)]-1},
\label{univ_r}
\end{equation}
which will have the same repercussions upon the later results. This allows us to intuit that there is a natural length scale to the Rutherford scattering, which is a notion that will only be reinforced further on, and repeatedly so.

\section{Fixed-target frame}
\label{fix_frame}

The target-relative position $\mathbf{r}$, as defined by (\ref{r}), immediately implies the fixed-target frame. In other words, for the absolute positions it holds by definition \mbox{$\mathbf{r}_\tar^{(\fix)}=\mathbf{0}$} and thus \mbox{$\mathbf{r}_\pro^{(\fix)}=\mathbf{r}$}. In order to determine the geometric shape of the shadow we pose the following question: under a particular angle $\theta$, which trajectory passes closest to the target? In other words, for a given $\theta$, which impact parameter $\R_0$ minimizes the distance $r(\theta;\R_0)$? The answer is, of course, to be found by finding the zero of the derivative in respect to $\R_0$:
\begin{equation}
\frac{\D r(\theta;\R_0)}{\D \R_0}\bigg|_{\tilde{\R}_0}=\frac{\tilde{\R}_0\sin\theta-2\X(1+\cos\theta)}{\tilde{\R}_0^3}r^2(\theta;\tilde{\R}_0)=0.
\label{mincond}
\end{equation}
This is satisfied by a vanishing numerator, yielding the sought impact parameter:
\begin{equation}
\tilde{\R}_0(\theta) =\frac{2\X}{\tan\frac{\theta}{2}}.
\label{tilde}
\end{equation}
Returning this value to (\ref{master}), we find that under an angle $\theta$ the trajectory with an impact parameter $\tilde{\R}_0$ comes closest to the target, at the distance:
\begin{equation}
r[\theta;\tilde{\R}_0(\theta)]=\frac{2\X}{\sin^2\frac{\theta}{2}}.
\label{rmin}
\end{equation}
Since $r[\theta;\tilde{\R}_0(\theta)]$ determines the shadow boundary, it is the solution to our problem: it represents the shadow equation in spherical coordinates. However, to make shadow shape more evident, we express its cylindrical coordinate $\rho$ (see figure~\ref{fig2}):
\begin{equation}
\rho(\theta)=r[\theta;\tilde{\R}_0(\theta)]\sin\theta=\frac{4\X}{\tan\frac{\theta}{2}},
\label{roth}
\end{equation}
as well as its $z$-coordinate:
\begin{equation}
z(\theta)=r[\theta;\tilde{\R}_0(\theta)]\cos\theta=2\X\left(\frac{1}{\tan^2\frac{\theta}{2}}-1\right).
\label{zeth}
\end{equation}
Eliminating the term $\tan(\theta/2)$ from previous two equations, the following connection is obtained:
\begin{equation}
z(\rho)=\frac{\rho^2}{8\X}-2\X,
\label{shadfix}
\end{equation}
which is the shadow equation in the cylindrical coordinates, and the main result of this paper. Evidently, in the fixed-target frame---as suggested by an example from figure~\ref{fig1}---all the projectile trajectories of a given energy form a paraboloidal shadow. As portended by (\ref{univ_r}), the shadow features a universal shape in a scaled coordinates \mbox{$\bar{z}=z/\X$} and \mbox{$\bar{\rho}=\rho/\X$}, such that: \mbox{$\bar{z}=\bar{\rho}^2/8-2$}, thus confirming the notion that there is a natural length scale to the Rutherford scattering. In addition, it is easily determined from the paraboloid's leading coefficient that the focal distance $f$ between the shadow focus and its vertex equals \mbox{$f=2\X$}, exactly corresponding to its free parameter. Therefore, in the fixed-target frame the target is precisely at the shadow focus!

\begin{figure}[t!]
\centering
\includegraphics[width=0.5\textwidth,keepaspectratio]{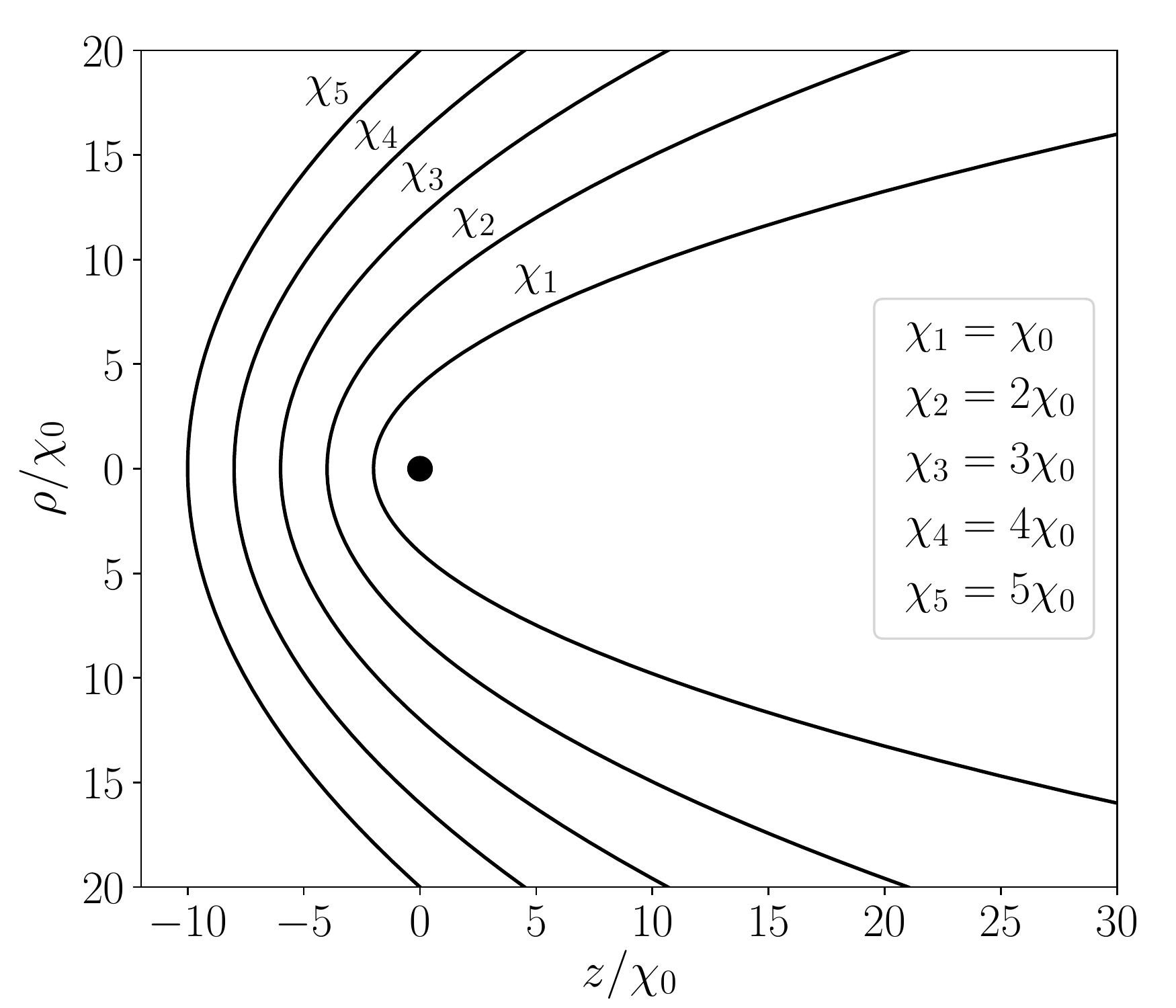}
\pull
\caption{Shadow examples in a fixed-target frame for several values of $\X$, with $\X_0$ as the arbitrary length scale. The charged target is shown by a central dot, which is precisely at the focus of each paraboloidal form. The shadow approaches the target and becomes stiffer as the initial energy of the projectile increases ($\X$ decreases).}
\label{fig3}
\end{figure}

Figure~\ref{fig3} shows shadow shapes for several arbitrary values $\X$, i.e. for several values of the initial projectile energy. From (\ref{x}) it is clear that the increase in the initial energy---i.e. in the initial relative speed $v_0$---means a decrease in $\X$. The shadows are perfectly in accord with expectations: not only do the projectiles of higher energy (lower $\X$) manage to come closer to the central target (as governed by the free parameter $-2\X$), but they are also less easily deflected than the projectiles of lower energy, meaning a stiffer paraboloid (as governed by the leading coefficient $1/8\X$).

Section~B of the Supplementary note offers some additional observations in regard to~(\ref{scatangle}), (\ref{tilde}) and (\ref{roth}).

\section{Center-of-mass frame}
\label{com_frame}

In order to make a transition from a fixed-target frame into any other frame, we invert the definitions of $\mathbf{R}$ and $\mathbf{r}$ from~(\ref{R}) and (\ref{r}), thus obtaining:
\begin{eqnarray}
&\mathbf{r}_\pro=\mathbf{R}+\frac{m_\tar}{m_\pro+m_\tar}\mathbf{r},\\
&\mathbf{r}_\tar=\mathbf{R}-\frac{m_\pro}{m_\pro+m_\tar}\mathbf{r}.
\end{eqnarray}
By definition, the center-of-mass position in the center-of-mass frame is the origin of the coordinate frame: \mbox{$\mathbf{R}^{(\cm)}=\mathbf{0}$}. Thus, introducing the shorthands:
\begin{equation}
\eta_{\pro,\tar}\equiv\frac{m_{\pro,\tar}}{m_\pro+m_\tar}
\end{equation}
we immediately obtain both the particle and target trajectories:
\begin{eqnarray}
&\mathbf{r}_\pro^{(\cm)}=\eta_\tar\mathbf{r},\\
&\mathbf{r}_\tar^{(\cm)}=-\eta_\pro\mathbf{r}.
\end{eqnarray}
As the projectile's position-vector in the center-of-mass frame is only scaled by the factor $\eta_\tar$ relative to the position in the fixed-target frame, the definition of the angle $\theta$ stays the same. The only effect upon the projectile trajectory is a decreased radial distance to the center of mass: \mbox{$r_\pro^{(\cm)}=\eta_\tar r$}, leading to the minimization condition:
\begin{equation}
\frac{\D r_\pro^{(\cm)}(\theta;\R_0)}{\D \R_0}\bigg|_{\tilde{\R}_0}=\eta_\tar\frac{\D r(\theta;\R_0)}{\D \R_0}\bigg|_{\tilde{\R}_0}=0.
\end{equation}
As the minimization procedure is unaffected in regard to (\ref{mincond}), the same minimizing value $\tilde{\R}_0(\theta)$ from (\ref{tilde}) is obtained. It is only that the minimum projectile distance from an origin of the center-of-mass frame is scaled by a factor $\eta_\tar$ when compared to that from (\ref{rmin}), with the same factor propagating into the cylindrical coordinates from~(\ref{roth}) and (\ref{zeth}), so that:
\begin{eqnarray}
&\rho_\pro(\theta)=\frac{4\eta_\tar\X}{\tan\frac{\theta}{2}},\\
&z_\pro(\theta)=2\eta_\tar\X\left(\frac{1}{\tan^2\frac{\theta}{2}}-1\right).
\end{eqnarray}
For brevity and clarity we have dropped the explicit frame designation (cm). Eliminating again the term $\tan(\theta/2)$ from previous two equations, we arrive at the shadow equation in the center-of-mass frame:
\begin{equation}
z_\pro(\rho_\pro)=\frac{\rho_\pro^2}{8\eta_\tar\X}-2\eta_\tar\X.
\label{compro}
\end{equation}
Since $\eta_t<1$, the paraboloid vertex is closer to the origin of the coordinate frame (as determined by the free parameter $-2\eta_\tar\X$), while the paraboloid shape is stiffer than in the fixed-target frame (as determined by the leading coefficient $1/8\eta_\tar\X$). Both effects are due to the fact that the origin is no longer the target itself, but rather the center-of-mass. Being somewhere in between the two particles, both the origin and the $z$-axis of the coordinate frame are \textit{at each point} along the particle trajectory brought closer to the projectile, when compared to the fixed-target frame, thus constricting the shadow profile.

In the center-of-mass frame the target is in motion, in an entirely symmetric manner as the projectile, so it also casts its own shadow. Its exact shape is easily deduced from the projectile shadow, as at any moment we may interchange the roles of the target and the projectile by a simple change in indices: \mbox{$\pro\leftrightarrow\tar$}. Additionally taking into account that the target shadow points at the opposite direction from the projectile shadow (along the negative direction of $z$-axis), we may immediately write:
\begin{equation}
z_\tar(\rho_\tar)=-\frac{\rho_\tar^2}{8\eta_\pro\X}+2\eta_\pro\X.
\label{comtar}
\end{equation}
Examining the focal distances $f_{\pro,\tar}$ of the projectile and target shadows from~(\ref{compro}) and (\ref{comtar}), we invariably find: $f_{\pro,\tar}=2\eta_{\tar,\pro}\X$, meaning that the focal points of both shadows are at the origin of the selected coordinate frame. Therefore, in the center-of-mass frame the two foci coincide, i.e. the same focus is shared between the two shadows!

If we were to examine the effect of varying masses upon the shadow form, it would not do to naively keep $\X$ constant, while varying only the ratios $\eta_{\tar,\pro}$ from products $\eta_{\tar,\pro}\X$ appearing in~(\ref{compro}) and (\ref{comtar}). This is because the term $\X$ itself, as defined by (\ref{x}), inherits the mass dependence via a reduced mass, so that the products $\eta_{\tar,\pro}\X$:
\begin{equation}
\eta_{\tar,\pro}\X\propto\frac{\eta_{\tar,\pro}}{\mu}=\frac{m_\pro+m_\tar}{m_\pro m_\tar}\frac{m_{\tar,\pro}}{m_\pro+m_\tar}=\frac{1}{m_{\pro,\tar}}
\label{chieta}
\end{equation}
are dependent only on the mass of a single particle. Therefore, in the center-of-mass frame the projectile and target shadows are determined solely by their own masses, being entirely independent of the other particle's mass. This needs to be held in mind, as it is in striking opposition with what the expression $\eta_{\tar,\pro}\X$ deceptively suggests: that the shadow form should not only be sensitive to both masses, but that it should also be more directly determined by the mass of the `wrong' particle.

In the sense of (\ref{chieta}), figure~\ref{fig3} may also be interpreted as a comparison of projectile shadows in the center-of-mass frame for varying projectile masses, if the labels $\chi_i$ are replaced by the projectile mass dependence \mbox{$m_i=m_0/i$} ($m_0$ being some arbitrary reference value).

Finally, it is again interesting to take note of the shadow form in the appropriately scaled coordinates. In fact, the projectile shadow from (\ref{compro}) takes on exactly the same universal form (\mbox{$\bar{z}=\bar{\rho}^2/8-2$}) of (\ref{shadfix}) when expressed in scaled coordinates $z_\pro/\eta_\tar\X$ and $\rho_\pro/\eta_\tar\X$. However, in order to reach the same form for the target shadow, its coordinates should be scaled by a different factor: $z_\tar/\eta_\pro\X$ and $\rho_\tar/\eta_\pro\X$. Therefore, scaling all coordinates by a unique factor $\X$ (or any constant, but parameter-invariant multiple of it) remains the most sensible choice. The price is that in the center-of-mass frame there is no parameter-independent universal form for both the projectile and the target simultaneously. Rather, with \mbox{$\bar{z}_{\pro,\tar}=z_{\pro,\tar}/\chi$} and \mbox{$\bar{\rho}_{\pro,\tar}=\rho_{\pro,\tar}/\chi$} we have to contend with two separate forms: \mbox{$\bar{z}_\pro=\bar{\rho}_\pro^2/8\eta_\tar-2\eta_\tar$} and \mbox{$\bar{z}_\tar=-\bar{\rho}_\tar^2/8\eta_\pro+2\eta_\pro$}, where the `most generalized' shadow shapes still depend on the relation between the two masses, but only on them. However, the advantage of this approach is that the length scale $\X$ is revealed not only as the most natural between the two particles (i.e. for both of them simultaneously), but also between multiple frames.

%, as the same one was used for the scaled version of (\ref{shadfix}).

\section{Quantum-mechanical case}
\label{quantum}

We present a short overview of the quantum-mechanical scattering and the appearance of the scattering shadow within such framework. A starting point is, of course, a Schr\"{o}dinger's equation for the joint particle-target system under a repulsive Coulomb interaction. After a typical separation of variables such that the motion of the center-of-mass is decoupled from the relative motion, an equation for the relative motion reads:
\begin{equation}
\left(-\frac{\hbar^2}{2\mu}\nabla^2+\frac{Z_\pro Z_\tar e^2}{4\pi\epsilon_0 r}\right)\psi_\mathbf{k}(\mathbf{r})=E_\mathbf{k}\psi_\mathbf{k}(\mathbf{r}),
\label{sch0}
\end{equation}
as a quantum-mechanical counterpart to the classical equation of motion~(\ref{eom}). Just like the Newton's equation, the Schr\"{o}dinger's equation for the relative motion features a reduced mass $\mu$. Since the relative vector $\mathbf{r}$ is still the same as in~(\ref{r})---its origin being at the target position---equation~(\ref{sch0}) describes a projectile in the fixed-target frame. In that, we have already parameterized both the wavefunction $\psi_\mathbf{k}(\mathbf{r})$ and the energy $E_\mathbf{k}$ of the relative motion by the wave vector $\mathbf{k}$ of the \textit{initial}, asymptotically free state of the system, described by the plane wave:
\begin{equation}
\lim_{\mathbf{k}\cdot\mathbf{r}\to-\infty}\psi_\mathbf{k}(\mathbf{r})\propto \mathrm{e}^{\mathrm{i}\mathbf{k}\cdot\mathbf{r}}
\label{asym}
\end{equation}
and serving as the boundary condition for solving~(\ref{sch0}). In order to establish the connection with the earlier classical treatment---in particular with the initial relative speed $v_0$ from~(\ref{v0}), which is otherwise, just like the trajectory, an ill-defined concept in quantum mechanics---we parameterize the initial wave vector as:
\begin{equation}
\mathbf{k}=\frac{\mu v_0}{\hbar}\hat{\mathbf{z}}.
\label{k}
\end{equation}
Introducing~(\ref{asym}) and (\ref{k}) into (\ref{sch0}) in the limit \mbox{$r\to\infty$}, a well known parameterization of energy remains: \mbox{$E_\mathbf{k}=\hbar k^2/2\mu=\mu v_0^2/2$}. With this, (\ref{sch0}) may be rewritten using the definition of $\X$ from (\ref{x}):
\begin{equation}
\left(\nabla^2+k^2-\frac{2\X k^2}{r}\right)\psi_\mathbf{k}(\mathbf{r})=0.
\label{sch}
\end{equation}
The solution to such Schr\"{o}dinger's equation, satisfying the boundary condition from~(\ref{asym}), is well known \cite{landau_quant}:
\begin{equation}
\psi_\mathbf{k}(\mathbf{r})=\mathrm{e}^{-\pi\X k/2}\Gamma(1+\mathrm{i}\X k)\mathrm{e}^{\mathrm{i}\mathbf{k}\cdot\mathbf{r}} M[-\mathrm{i}\X k, 1, \mathrm{i}(kr-\mathbf{k}\cdot\mathbf{r})],
\label{psi}
\end{equation}
with $\Gamma$ as the conventionally defined gamma-function and $M$ as the Kummer's confluent hypergeometric function, otherwise denoted as $_1F_1$. The wavefunctions from~(\ref{psi}) are normalized such that: \mbox{$\int \psi_\mathbf{k}^*(\mathbf{r})\psi_{\mathbf{k}'}(\mathbf{r}) \D V =(2\pi)^3 \delta(\mathbf{k}-\mathbf{k}')$}.

As opposed to classical mechanics, where all trajectories are strictly excluded from the shadow zone, in quantum mechanics we would always expect the wavefunction tunneling into this classically forbidden region of space. A question naturally arises if the wavefunction exhibits any recognizable features at all, that would allow us to identify the appearance of the classical shadow. In general case, its precise position could hardly be pinpointed from the Coulomb continuum wavefunction, as the quantum shadow is diffuse. However, we can make the some observations in the opposite direction: knowing the classical shadow, we can analyze the wavefunction behavior in its vicinity. Burgd\"{o}rfer notices: `\textit{A (smoothed) caustic appears also in the corresponding quantum scattering wavefunction as an (anti) nodal surface. The locus of the caustic is, in fact, most conveniently derived from the nodal structure of Coulomb continuum wavefunctions. Nodal surfaces are given by the argument of the hypergeometric function (...)}' (quotation from \cite{shadow2}). In this rather ingenious insight, we only caution against the use of the term `(anti) nodal', as it might suggest that the shadow appears at some extremum related to the wavefunction---presumably the extermum of modulus $|\psi_\mathbf{k}(\mathbf{r})|$---which we will soon disprove. The more appropriate term would be `level surfaces' (`equipotentials'), which are indeed defined by the constancy of the argument of the confluent hypergeometric function from~(\ref{psi}):
\begin{equation}
kr-\mathbf{k}\cdot\mathbf{r}=C.
\label{const}
\end{equation}
For $\mathbf{k}$ defined as in~(\ref{k}), (\ref{const}) reduces to \mbox{$k(\sqrt{z^2+\rho^2}-z)=C$}. Solving for $z$ yields:
\begin{equation}
z(\rho)=\frac{k}{2C}\rho^2-\frac{C}{2k}=\frac{\rho^2}{8(C/4k)}-2(C/4k)
\label{level}
\end{equation}
for the shape of level surfaces of $|\psi_\mathbf{k}(\mathbf{r})|$ (but not of $\psi_\mathbf{k}(\mathbf{r})$ itself, due to the extra $\mathrm{e}^{\mathrm{i}\mathbf{k}\cdot\mathbf{r}}$ factor). This has the same form as~(\ref{shadfix}), allowing to determine the shadow-related value of constant $C$ as:
\begin{equation}
C_\mathrm{shadow}=4\X k
\label{c_shad}
\end{equation}
and, indeed, to recognize the classical scattering shadow as a particular level surface in the quantum-mechanical probability density of the incoming projectile.

\begin{figure}[t!]
\centering
\includegraphics[width=0.5\textwidth,keepaspectratio]{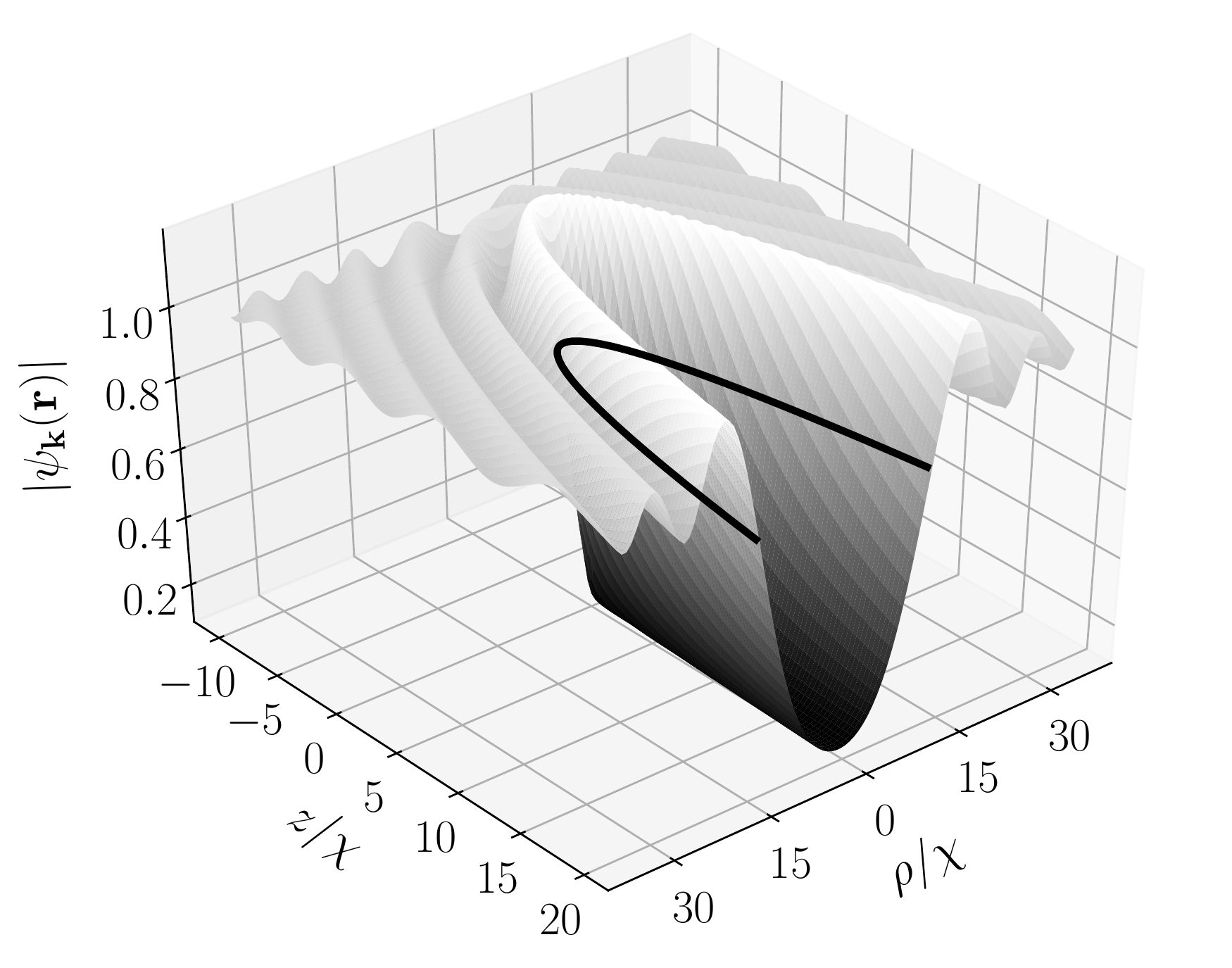}
\pull
\caption{Modulus of the projectile's Coulomb continuum wavefunction for a repulsive Rutherford scattering in the fixed-target frame, in a plane containing the $z$-axis, with the target at \mbox{$\mathbf{r}=\mathbf{0}$}. The relevant wavefunction features are governed by the confluent hypergeometric function \mbox{$M[-\mathrm{i}\X k, 1, \mathrm{i}\X k(\sqrt{\bar{\rho}^2+\bar{z}^2}-\bar{z})]$}, with \mbox{$\bar{z}=z/\X$} and \mbox{$\bar{\rho}=\rho/\X$}, and here selected \mbox{$\X k=1$}. The thick black line shows the classical shadow caustic, beyond which the wavefunction clearly exhibits a quantum-mechanical tunneling.}
\label{fig4}
\end{figure}

Figure~\ref{fig4} shows an example of the modulus $|\psi_\mathbf{k}(\mathbf{r})|$ of the wavefunction from~(\ref{psi}) for \mbox{$\X k=1$}, in a plane containing the $z$-axis, where the target rests at \mbox{$\mathbf{r}=\mathbf{0}$}. One can readily appreciate by eye the fact that the level surfaces are parabolic, as shown by~(\ref{level}). The thick black line indicates the level surface corresponding to the shadow caustic---i.e. where, along the wavefunction, the classical shadow appears---clearly proving that it is not related to any extremal (antinodal) surface. The portion of the wavefunction bounded by this caustic (below the thick black line) shows a clear case of the quantum-mechanical tunneling into the classically forbidden zone.

The fact that the classical scattering shadow may indeed be identified within the quantum-mechanical description indicates that we might again perform the appropriate coordinate scaling---such that \mbox{$\bar{z}=z/\X$} and \mbox{$\bar{r}=r/\X$}---and express the wavefunction as:
\begin{equation}
\psi_\mathbf{k}(\mathbf{r})=\mathrm{e}^{-\pi\X k/2}\Gamma(1+\mathrm{i}\X k)\mathrm{e}^{\mathrm{i}\X k\bar{z}} M[-\mathrm{i}\X k, 1, \mathrm{i}\X k(\bar{r}-\bar{z})],
\label{psi2d}
\end{equation}
where we used, for simplicity, the convention $\mathbf{k}=k\hat{\mathbf{z}}$ from~(\ref{k}). This reveals that, while the shape of the scattering shadow still remains scale-invariant, the details of the wavefunction still depend on $k$, but in such way that---alongside the length scale $\X$---there appears another, dimensionless scale $\X k$, otherwise known as Sommerfeld parameter. But how can that be, considering that in the classical mechanics \textit{all} the spatial aspects of the Coulomb trajectories from~(\ref{univ_r}) are scaled only by $\X$? How can another scale be admitted in quantum-mechanics, since---by the correspondence principle---at some point both the classical and quantum description must coincide? The answer lies in the temporal aspects of the scattering, of which the purely geometrical expression~(\ref{univ_r}) has been devoided. For the same spatial scaling $\X$, the projectile speed---entering $k$ through~(\ref{k})---may still be varied. Thus, the time the projectile spends in a given portion of space still depends on its speed. In consequence, so do the details of the wavefunction, governing the probability density $|\psi_\mathbf{k}(\mathbf{r})|^2$ of finding the projectile at a given point. This probabilistic interpretation, in combination with correspondence principle applied to the projectile trajectories displayed in figure~\ref{fig1}, also allows us to understand why the highest antinode in $|\psi_\mathbf{k}(\mathbf{r})|$ appears just before the shadow, prior to tunneling. It is for two reasons: for $z/\X\lesssim -1$ the projectiles are slowest just around the scattering shadow (see Section~C of the Supplementary note), while for $z/\X\gtrsim -1$ the trajectories pile around the shadow caustic (see figure~\ref{fig1}), both effects increasing the probability of finding the projectiles at the edge of the classically forbidden zone.

\begin{figure}[t!]
\centering
\includegraphics[width=0.5\textwidth,keepaspectratio]{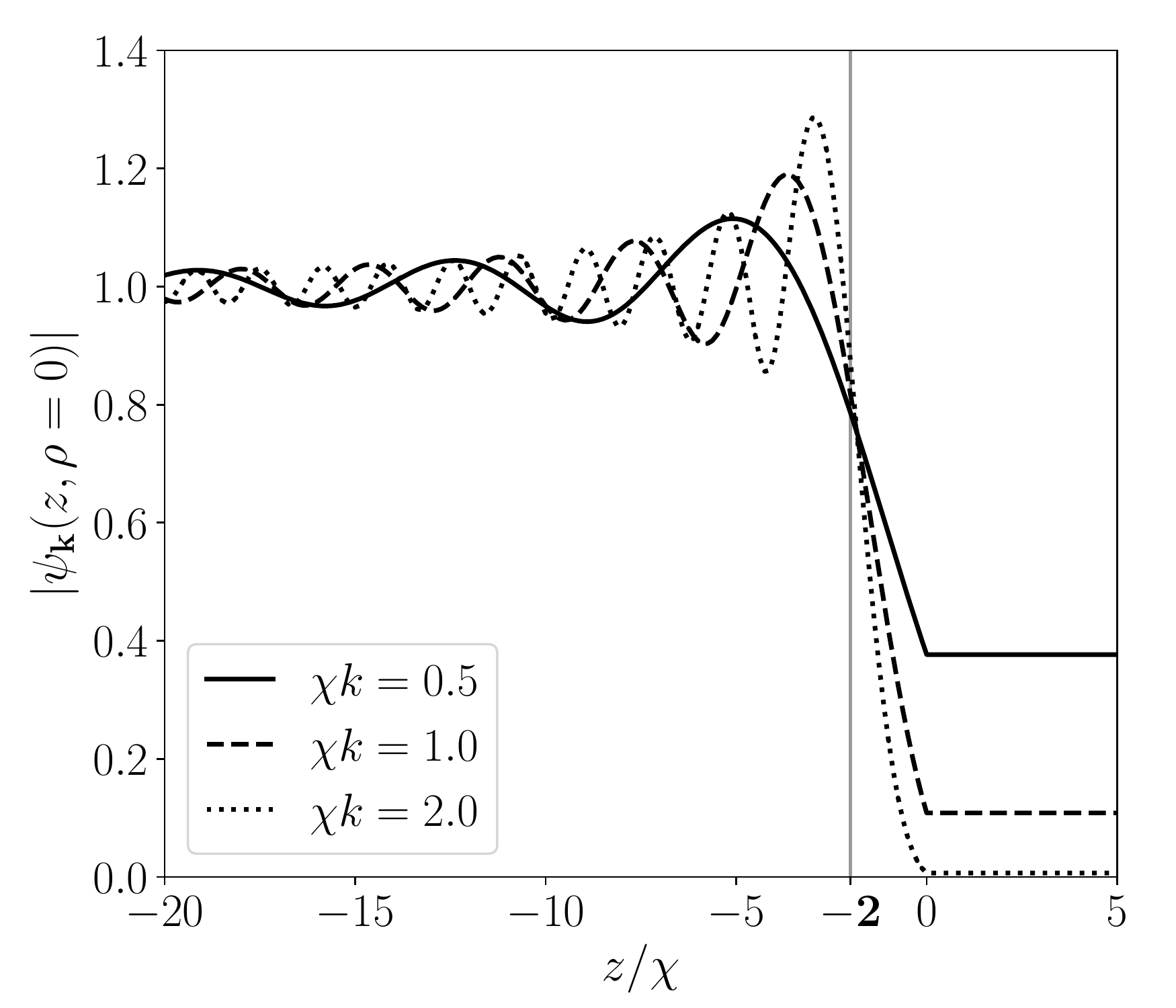}
\pull
\caption{Modulus of the Coulomb continuum wavefunction for the repulsive Rutherford scattering along the axis \mbox{$\rho=0$}, for different values of the Sommerfeld parameter $\X k$. The target is at \textbf{$z=0$}. The vertex of the classical shadow caustic is indicated by the vertical line at \textbf{$z/\X=-2$}.}
\label{fig5}
\end{figure}

Wavefunction dependence upon $\X k$ is further exemplified by figure~\ref{fig5}, showing the modulus $|\psi_\mathbf{k}(\mathbf{r})|$ along the central axis \mbox{$\rho=0$}, passing through a target at \textbf{$z=0$}, for different values of $\X k$. The position of the classical shadow caustic---now corresponding to its vertex---is shown by the vertical line. There is no single point at which all the wavefunctions intersect, as might be falsely inferred from this specific display.

Figure~\ref{fig5} allows us to make some additional interesting observations. Since at \mbox{$\rho=0$} it holds: \mbox{$r=|z|$}, for \mbox{$z\ge0$} the confluent hypergeometric function from~(\ref{psi2d}) reduces to \mbox{$M(-\mathrm{i}\X k, 1, 0)=1$}, so $|\psi_\mathbf{k}(\mathbf{r})|$ is indeed constant there, as suggested by the figure. At the first glance it might be confusing why this value is not 0 at $z=0$, where the target lies. In other words, how can the Schr\"{o}dinger's equation from~(\ref{sch0}) or (\ref{sch}) be satisfied by a nonvanishing wavefunction at the point where the repulsive potential diverges? As figure~\ref{fig5} shows, the wavefunction  at \mbox{$z=0$}, while continuous, is not smooth---its gradient $\nabla\psi_\mathbf{k}(\mathbf{r})$ is discontinuous, so that its Laplacian $\nabla^2\psi_\mathbf{k}(\mathbf{r})$ diverges, canceling the divergence from the potential energy in the Schr\"{o}dinger's equation. Finally, figure~\ref{fig5} also indicates that by increasing $\X k$ the wavefunction behavior around the shadow caustic becomes sharper and sharper. This suggests that if one were to investigate the limit \mbox{$\X k\to\infty$} of a very strong repulsion and/or a very slow projectile---\mbox{$\lim_{\X k\to \infty}|\psi_\mathbf{k}(\mathbf{r})|\propto \lim_{\X k\to \infty}|M(-\mathrm{i}\X k, 1,\mathrm{i}\X k\xi)|$} with, in our case, \mbox{$\xi=(r-z)/\X$}---one would expect a sharp drop at $\xi=4$, in accordance with~(\ref{c_shad}). This specific result could then be extended to a case of finite $\X k$ and taken as an agreed-upon value determining the shadow caustic even in the case of a diffuse shadow. This is how the scattering shadow can be determined self-consistently from the wavefunction itself, without reference to the classical mechanics.

\section{Conclusion}
\label{conclusion}

We have explored the geometry of the repulsive Rutherford scattering, finding the exact shape of the scattering shadow in the fixed-target and the center-of-mass frame. In both frames the projectile shadow has a simple paraboloidal shape. The difference between frames is, of course, reflected in different values of the paraboloids' coefficients, i.e. in their stiffness and the distance of their vertices from the origin of the coordinate frame. In that, the projectile shadow in the center-of-mass frame is stiffer and closer to the origin than its counterpart from the fixed-target frame. Since the motion of the target in the center-of-mass frame is---in mathematical form---symmetrical to the motion of the projectile, the target also casts a paraboloidal shadow in the same frame, such that the two shadows intersect. It was found that the target is precisely at the shadow focus in the fixed-target frame, while in the center-of-mass frame the focal points of the projectile and target shadow coincide with the center of mass itself. A somewhat surprising finding is that the shadow parameters in the center-of-mass frame depend only on the mass of the particle casting the shadow, rather than the ratio of masses as might at first be expected. The Rutherford scattering is revealed to feature a natural length scale $\X$, determined by the physical parameters of the system. A quantum-mechanical treatment of the repulsive Rutherford scattering was addressed and the scattering shadow was also observed appearing in a Coulomb continuum wavefunction. Unlike the sharply defined classical shadow, quantum mechanics yields a diffuse shadow caustic, due to the wavefunction tunneling into a classically forbidden zone. Alongside a length scale $\X$, in quantum mechanics another relevant scale appears: a dimensionless Sommerfeld parameter $\X k$. A sharp shadow caustic is recovered in the limit \mbox{$\X k\to \infty$}. The transition of the scattering shadow to the laboratory frame---wherein the target is at rest only at the initial moment, subsequently being recoiled by the approaching projectile---is much more involved and will be the subject of the future work.

%\clearpage

\begin{acknowledgments}

We are grateful to Ivica Smoli\'{c} and Kre\v{s}imir Dekani\'{c} for the useful discussions and for the help in tracking down the relevant literature.

\end{acknowledgments}

%===========================================================================================

\clearpage

\thispagestyle{empty}

\setcounter{page}{1}
\setcounter{section}{0}

\renewcommand{\thesection}{\Alph{section}}
\numberwithin{equation}{section}
\numberwithin{figure}{section}
\numberwithin{table}{section}

\renewcommand{\theequation}{\thesection\arabic{equation}}
\renewcommand{\thefigure}{\thesection\arabic{figure}}

\renewcommand{\thepage}{S\arabic{page}}  

%\renewcommand{\bibnumfmt}[1]{[#1]}
%\renewcommand{\citenumfont}[1]{#1}

%ODAVDE BRISAT

\begin{comment}

\title{{\LARGE Supplementary note}\\ \vspace*{3mm} A shadow of the repulsive Rutherford scattering in the fixed-target and the center-of-mass frame}

\author{Petar \v{Z}ugec$^1$}
\email{pzugec@phy.hr}
\author{Ivan Topi\'{c}$^2$}

\affiliation{$^1$Department of Physics, Faculty of Science, University of Zagreb, Zagreb, Croatia}
\affiliation{$^2$Archdiocesan Classical Gymnasium, Zagreb, Croatia}

%\date{\today}

\begin{abstract}

This note presents the supplementary material to the main paper. The references to figures and equations not starting with the alphabetical character---such as~(1)---refer to those from the main paper, while those starting with the appropriate letter---e.g.~(A1)---refer to those from this note.

\end{abstract}

\maketitle 

\end{comment}

%DOVDE BRISAT

\onecolumngrid
\begin{center}

\textbf{\LARGE Supplementary note}\\[.5cm]

\textbf{\large A shadow of the repulsive Rutherford scattering in the fixed-target and the center-of-mass frame}\\[.5cm]

Petar \v{Z}ugec$^{1*}$ and Ivan Topi\'{c}$^2$\\[.1cm]
{\small
{\itshape
$^1$Department of Physics, Faculty of Science, University of Zagreb, Zagreb, Croatia\\
$^2$Archdiocesan Classical Gymnasium, Zagreb, Croatia\\}
$^*$Electronic address: pzugec@phy.hr\\[.5cm]
%(Dated: \today)\\[.5cm]
}

%{438pt}
\begin{minipage}{400pt}
\small
This note presents the supplementary material to the main paper. The references to figures and equations not starting with the alphabetical character---such as~(1)---refer to those from the main paper, while those starting with the appropriate letter---e.g.~(A1)---refer to those from this note.\\
\end{minipage}

\end{center}

\twocolumngrid

\section{Coulomb trajectories derivation}

We start with the equation of motion:
\begin{equation}
\ddot{\mathbf{r}}=\frac{Z_\pro Z_\tar e^2}{4\pi\epsilon_0\mu}\frac{\hat{\mathbf{r}}}{r^2},
\label{eomx}
\end{equation}
introduced in~(\ref{eom}). In the central-force field the total angular momentum $\mathbf{L}$ of the system is conserved---as the torque $\mathbf{T}$ vanishes due to the collinearity of the position and force vectors \mbox{($\mathbf{T}=\D\mathbf{L}/\D t=\mathbf{r}\times\mathbf{F}_{\tar\rightarrow\pro}=\mathbf{0}$)}---which means that the motion of the system is constrained to a single plane. By selecting the plane with the constant azimuthal coordinate ($\varphi=\mathrm{const.}$), the acceleration term $\ddot{\mathbf{r}}$ expressed in spherical coordinates takes the form:
\begin{equation}
\ddot{\mathbf{r}}=(\ddot{r}-r\dot{\theta}^2)\hat{\mathbf{r}}+\frac{1}{r}\frac{\D(r^2\dot{\theta})}{\D t}\hat{\boldsymbol{\theta}},
\label{sph}
\end{equation}
where the spherical (spatial) coordinates $r$ and $\theta$ now act as the polar (planar) coordinates within the plane defined by $\varphi=\mathrm{const.}$ (One can obtain a sense of the relevant geometric parameters from figure~\ref{fig2}.) By comparing~(\ref{eomx})~and~(\ref{sph}), two equations of motion are readily obtained. The first follows from the vanishing derivative term related to the unit $\hat{\boldsymbol{\theta}}$-direction, meaning that the $r^2\dot{\theta}$ term must be constant. As we will assume that the projectile is put into motion from infinity on the negative side of the $z$-axis (i.e. from $\theta_0=\pi$; see figure~\ref{fig2}), the angular change rate $\dot{\theta}$ will be negative. We choose to parameterize the associated constant as \mbox{$r^2\dot{\theta}=-\ell$}, with $\ell$ being positive. Therefore:
\begin{equation}
\dot{\theta}=-\frac{\ell}{r^2},
\label{doth}
\end{equation}
where the value of $\ell$ will be determined later, from the initial conditions.

The second equation of motion follows from equating the terms by the unit $\hat{\mathbf{r}}$-direction in~(\ref{eomx})~and~(\ref{sph}):
\begin{equation}
\ddot{r}-r\dot{\theta}^2=\frac{Z_\pro Z_\tar e^2}{4\pi\epsilon_0\mu}\frac{1}{r^2}.
\label{rad}
\end{equation}\\\\
This equation is commonly solved by introducing the substitution $u=1/r$. Together with (\ref{doth}), we use it to first translate the time derivative into the associated angular derivative:
\begin{equation}
\frac{\D}{\D t}=\frac{\D\theta}{\D t}\frac{\D}{\D\theta}=-\frac{\ell}{r^2}\frac{\D}{\D\theta}=-\ell u^2\frac{\D}{\D\theta}.
\label{ddt}
\end{equation}
From here it follows:
\begin{equation}
\ddot{r}=\frac{\D}{\D t}\left(\frac{\D}{\D t}\frac{1}{u}\right)=-\ell u^2\frac{\D}{\D\theta}\left(-\ell u^2\frac{\D}{\D\theta}\frac{1}{u}\right)=-\ell^2 u^2\frac{\D^2u}{\D\theta^2}.
\label{ddor}
\end{equation}
Plugging~(\ref{doth})~and~(\ref{ddor}) back into (\ref{rad}) leaves us with:
\begin{equation}
\frac{\D^2u}{\D\theta^2}+u=-\frac{\kappa}{\ell^2},
\label{binet}
\end{equation}
where we have temporarily introduced the constant \mbox{$\kappa\equiv Z_\pro Z_\tar e^2/4\pi\epsilon_0\mu$} defined by the intrinsic system parameters (charges and masses), without depending on the initial conditions. Equation~(\ref{binet}) is a well known Binet equation. Having a familiar form of the shifted harmonic oscillator equation, its solution is easily found as:
\begin{equation}
u(\theta)=\U \cos(\theta-\T)-\frac{\kappa}{\ell^2},
\label{solu}
\end{equation}
with the constants $\U$ and $\T$, together with $\ell$, to be determined from the initial conditions.

In parameterizing the projectile trajectory we will make use of the cylindrical coordinates $\rho$ and $z$, alongside their spherical counterparts $r$ and $\theta$ used up to this point. Figure~\ref{fig2} clearly illustrates their relation. In that, the direction of the $z$-axis corresponds to the projectile's initial direction of motion (i.e. its initial velocity). It always holds: \mbox{$\mathbf{r}=r\hat{\mathbf{r}}=\rho\hat{\boldsymbol{\rho}}+z\hat{\mathbf{z}}$}, regardless of the specific functional dependency of the coordinates and unit directions, whether it be angular or temporal. Assuming that the projectile has been put into motion as a free particle of initial speed $v_0$ and with the impact parameter  $\R_0$, from the negative side of the $z$-axis, at the infinite distance from the target ($\theta_0=\pi$), we can write:
\begin{align}
&\mathbf{r}(\theta_0=\pi)=\R_0\hat{\boldsymbol{\rho}}+\Big(\lim_{z_0\to-\infty}z_0\Big)\hat{\mathbf{z}},\\
&\dot{\mathbf{r}}(\theta_0=\pi)=v_0\hat{\mathbf{z}}.
\end{align}\clearpage

\noindent
From $\mathbf{r}=r\hat{\mathbf{r}}$ and \mbox{$\dot{\mathbf{r}}=\dot{r}\hat{\mathbf{r}}+r\dot{\theta}\hat{\boldsymbol{\theta}}$} it is easy to show that the constant $\ell$, as we have defined it, equals to\footnote{
It may be shown that the constant $\ell$ is related to the total angular momentum $\mathbf{L}^{(\cm)}$ in the center-of-mass frame: \mbox{$\ell=|\mathbf{L}^{(\cm)}|/\mu$}. However, this parametrization is of limited use to this work.
}:
\begin{equation}
\ell=|r^2\dot{\theta}|=|\mathbf{r}\times\dot{\mathbf{r}}|=\R_0v_0.
\label{ell}
\end{equation}
Applying the initial position condition \mbox{$r(\theta_0=\pi)=\infty$}---that is \mbox{$u(\theta_0=\pi)=0$}---to (\ref{solu}), it follows:
\begin{equation}
\U \cos(\pi-\T)-\frac{\kappa}{\ell^2}=0.
\label{inpos}
\end{equation}
Applying (\ref{ddt}), it is easily shown that: \mbox{$\D r(\theta)/\D t=\ell\: \D u(\theta)/\D\theta=-\U \ell\sin(\theta-\Theta)$}, so the initial speed condition \mbox{$\dot{r}(\theta_0=\pi)=-v_0$} (negative sign due to the initial reduction of the radial distance, as the projectile approaches the target) translates into:
\begin{equation}
\U \ell \sin(\pi-\T)=v_0
\label{inspd}
\end{equation}
Using (\ref{ell}),~(\ref{inpos})~and~(\ref{inspd}) are to be solved for $\U$ and $\T$, yielding:
\begin{align}
&\U=\frac{1}{\R_0}\sqrt{1+\left(\frac{\kappa}{\R_0 v_0^2}\right)^2},
\label{uu}\\
&\T=\frac{\pi}{2}+\atan\left(\frac{\kappa}{\R_0 v_0^2}\right).
\label{tt}
\end{align}
As we will have to analyze the solution dependence on the impact parameter $\R_0$, we define the following term absorbing all the parameters save $\R_0$ itself\footnote{
In a fixed-target frame the initial kinetic energy \mbox{$E_0=m_\pro v_0^2/2$} of the projectile is as good parameter as the initial relative speed $v_0$, allowing the parameter $\X$ from (\ref{xap}) to be expressed as:
\begin{equation*}
\X^{(\fix)}=\frac{Z_\pro Z_\tar e^2m_\pro}{8\pi\epsilon_0\mu E_0}.
\end{equation*}
However, in any other frame (moving relative to the target) the parametrization by energy becomes cumbersome, as it transforms between the frames, while the initial projectile energy does not correspond any more to the total energy of the system. On the other hand, the initial relative speed $v_0$ remains the same in all frames, providing a frame-independent parametrization of $\X$.
}:
\begin{equation}
\X\equiv\frac{\kappa}{v_0^2}=\frac{Z_\pro Z_\tar e^2}{4\pi\epsilon_0\mu v_0^2}.
\label{xap}
\end{equation}
Plugging~(\ref{ell}), (\ref{uu}) and (\ref{tt}) into (\ref{solu}) we may write the final solution (recall that $r=1/u$) as:
\begin{equation}
r(\theta;\R_0)=\frac{\R_0^2}{\sqrt{\X^2+\R_0^2}\sin[\theta-\atan(\X/\R_0)]-\X}.
\label{masterap}
\end{equation}
Alternative expressions for $r(\theta;\R_0)$ include:
\begin{align}
\begin{split}
r(\theta;\R_0)&=\frac{\R_0^2}{\R_0\sin\theta-\X(1+\cos\theta)}\\
&=\frac{\R_0^2}{2\cos^2\frac{\theta}{2}(\R_0\tan\frac{\theta}{2}-\X)},
\end{split}
\end{align}
having been obtained by a simple use of the trigonometric identities.

\section{Some geometric observations}

\begin{figure}[h!]
\centering
\includegraphics[width=0.5\textwidth,keepaspectratio]{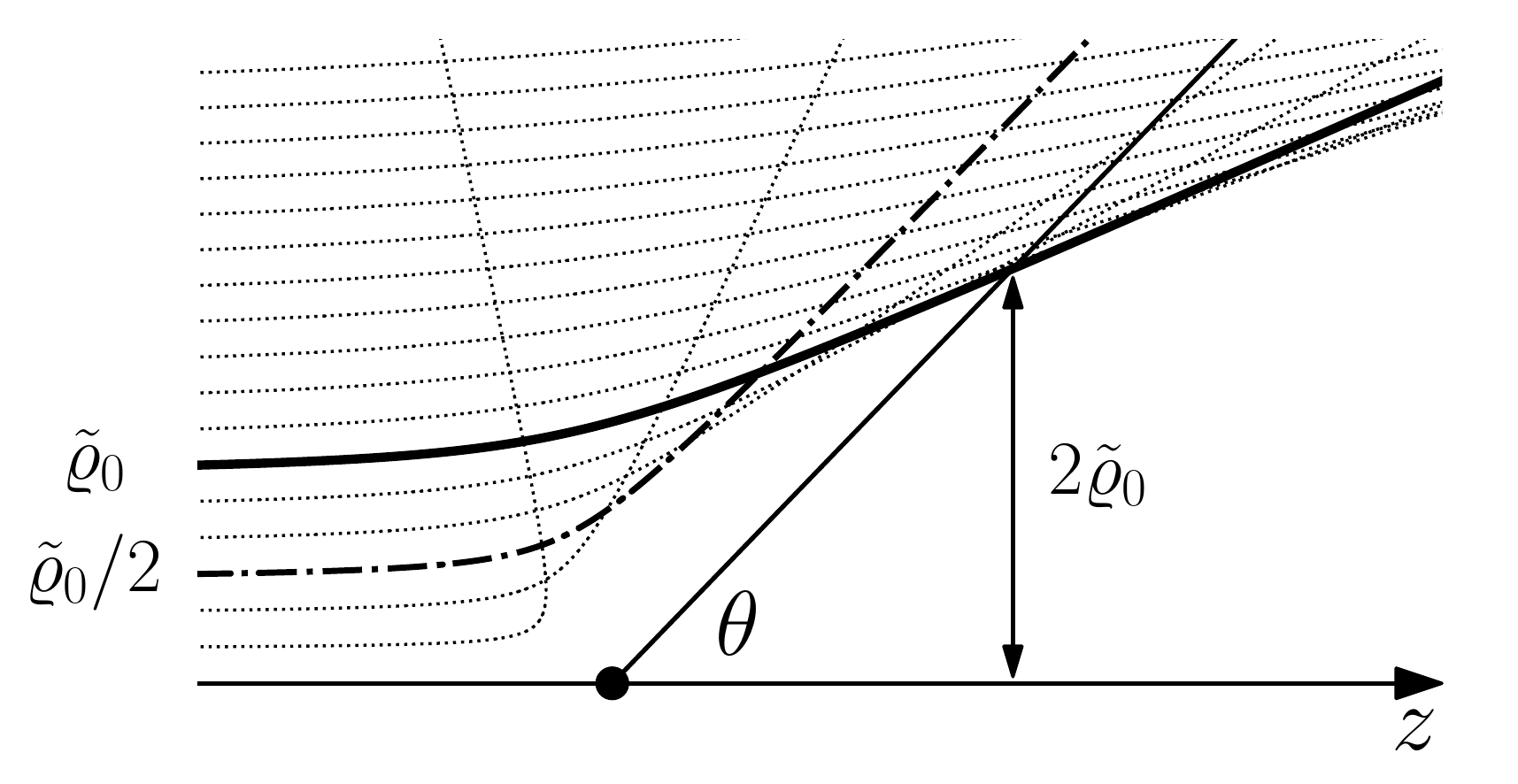}
\pull
\caption{Geometric relation between several notable parameters. The trajectory that minimizes the projectile-target distance under a given angle $\theta$ (full thick line) starts from an impact parameter $\tilde{\R}_0$ which is double the value of the lowest one ($\tilde{\R}_0/2$; thick dashed trajectory) required to even reach the same angle. At the distance-minimization point the corresponding trajectory has doubled its radial distance from the $z$-axis ($2\tilde{\R}_0$). Thin dashed trajectories show several examples that either do not minimize the projectile-target distance under the selected angle or do not even reach it.\pullc}
\label{figx}
\end{figure}

In regard to (\ref{scatangle}), as well as in~(\ref{tilde}) and (\ref{roth}) several very similar quantities have appeared:
\begin{equation*}
\frac{\X}{\tan\frac{\theta}{2}} \quad\mathrm{and}\quad \frac{2\X}{\tan\frac{\theta}{2}} \quad\mathrm{and}\quad \frac{4\X}{\tan\frac{\theta}{2}}.
\end{equation*}
Taking the middle one---corresponding to the term $\tilde{\R}_0$ from (\ref{tilde})---as a reference, we are dealing the equivalent sequence: \mbox{$\tilde{\R}_0/2$, $\tilde{\R}_0$, $2\tilde{\R}_0$}. Figure~\ref{figx} shows the geometric meaning of these values. The reference $\tilde{\R}_0$ is the impact parameter minimizing the projectile-target distance under a given angle $\theta$. \textit{At this point} the projectile moving along a distance-minimizing trajectory has doubled its radial distance from the $z$-axis: from the starting $\tilde{\R}_0$ to $2\tilde{\R}_0$. Finally, only the trajectories distant enough from the target---those with impact parameter greater than half of the value that minimizes the projectile-target distance under a given angle (\mbox{$\R_0>\tilde{\R}_0/2$})---can even reach the same angle, while the rest are scattered before this point.

\section{The closest approach}

The first, intuitive thought that may come to mind in attempting to obtain the shadow shape is that it might be determined by the trajectories' points of the closest approach, that are---unlike the shadow itself---often quoted in the literature. However, after a brief contemplation one is quickly disabused of that notion, as one realizes that what is commonly quoted as the `point of the closest approach' refers to the closest approach point \textit{from a given trajectory}. On the other hand, the shadow itself is determined by the closest approach trajectories \textit{from all possible trajectories}. It is instructive to examine this difference in detail and to determine the actual geometric place of the points of the closest approach. We carry out this analysis only in the fixed-target frame.

\begin{figure}[t!]
\centering
\includegraphics[width=0.5\textwidth,keepaspectratio]{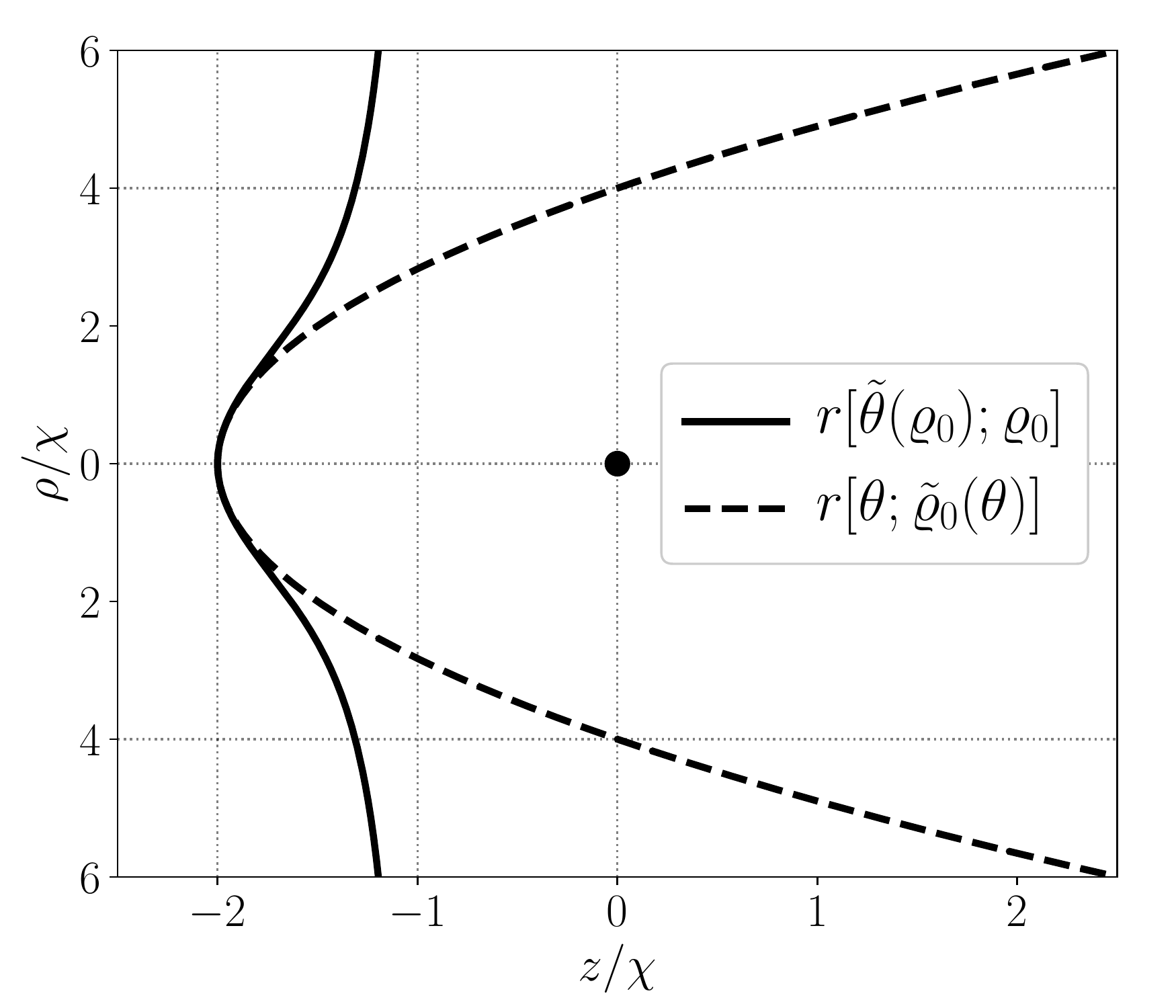}
\pull
\caption{Universal shape of the geometric place of the closest approach points (full line) compared to the universal shape of the scattering shadow (dashed line) in a fixed-target frame (target is the central dot). The closest approach curve has an asymptote at \mbox{$z/\X=-1$}, so that it is fully contained within \mbox{$-2\le z/\X<-1$}. While $\tilde{\theta}(\R_0)$ minimizes the target-projectile distance \textit{for a given projectile trajectory} (of given impact parameter $\R_0$), $\tilde{\R}_0(\theta)$ minimizes \textit{among all possible trajectories} the target-projectile distance under a given angle $\theta$.\pullc}
\label{figA1}
\end{figure}

For a given trajectory, i.e. a given impact parameter $\R_0$, the closest approach is determined by minimizing the target-projectile distance from (\ref{masterap}) in respect to $\theta$. It is easily done just by observing that the expression is minimal when the denominator is maximal. This is fulfilled when the sine term itself is maximized, i.e. when its argument equals $\pi/2$, immediately yielding an angle:
\begin{equation}
\tilde{\theta}(\R_0)=\frac{\pi}{2}+\atan\frac{\X}{\R_0},
\label{tilde_th}
\end{equation}
under which the trajectory comes closest to the target\footnote{
Relative to the initial angle \mbox{$\theta_0=\pi$}---and due to the symmetric shape of the hyperbole---the scattering angle $\vartheta$ from (\ref{scatangle}) is always double the angle $\tilde{\theta}$ of the closest approach from (\ref{tilde_th}):
\begin{equation*}
\vartheta-\theta_0=2(\tilde{\theta}-\theta_0).
\end{equation*}
In other words, the hyperbole is symmetric around $\tilde{\theta}$.
}, at a distance:
\begin{equation}
r[\tilde{\theta}(\R_0);\R_0]=\frac{\R_0^2}{\sqrt{\X^2+\R_0^2}-\X}.
\label{rmin_rho}
\end{equation}
If we invert (\ref{tilde_th}) in order to find the impact parameter corresponding to a particular minimizing angle $\tilde{\theta}$: \mbox{$\R_0(\tilde{\theta})=-\X\tan\tilde{\theta}$}, we may express the minimized distance from (\ref{rmin_rho}) as a function of the same angle\footnote{
In applying the identity \mbox{$\sqrt{1+\tan^2\tilde{\theta}}=1/|\cos\tilde{\theta}|$} to arrive at (\ref{rmin_th}), one needs to take: \mbox{$|\cos\tilde{\theta}|=-\cos\tilde{\theta}$} as for \mbox{$\pi<\tilde{\theta}<\pi/2$} the cosine is negative.
}:
\begin{equation}
r[\tilde{\theta};\R_0(\tilde{\theta})]=\X\frac{\cos\tilde{\theta}-1}{\cos\tilde{\theta}},
\label{rmin_th}
\end{equation}
which is the sought geometric place of the all points of the closest approach, in spherical coordinates. Already from the form of this expression one can conclude that it can not possibly describe the shadow of the Rutherford scattering, as this expression for the absolute distance is positive only for \mbox{$\tilde{\theta}>\pi/2$}, while the shadow is expected to cover also the forward angles (\mbox{$\theta<\pi/2$}; see figure~\ref{fig1}). Employing again the coordinate transformations \mbox{$z(\tilde{\theta})=r[\tilde{\theta};\R_0(\tilde{\theta})]\cos\tilde{\theta}$} and \mbox{$\rho(\tilde{\theta})=r[\tilde{\theta};\R_0(\tilde{\theta})]\sin\tilde{\theta}$}, while eliminating the term $\cos\tilde{\theta}$ from $z(\tilde{\theta})$, one obtains the explicit shape of the curve from (\ref{rmin_th}) in cylindrical coordinates:
\begin{equation}
\rho(z)=\frac{z\sqrt{-z(z+2\X)}}{z+\X}.
\label{rhomin}
\end{equation}
It is to be noted yet again that in scaled coordinates $\bar{\rho}=\rho/\X$ and $\bar{z}=z/\X$ this expression also has a universal form: \mbox{$\bar{\rho}=\bar{z}\sqrt{-\bar{z}(\bar{z}+2)}/(\bar{z}+1)$}. While (\ref{rmin_th}) is positive for $\tilde{\theta}>\pi/2$---suggesting that the closest approach curve might be defined up to \mbox{$z_\lm=0$}---both the dependence \mbox{$z(\tilde{\theta})=\X(\cos\tilde{\theta}-1)$} and the fact that (\ref{rhomin}) is nonnegative only for \mbox{$-2\X\le z<-\X$} reveal that the curve's asymptote is actually at \mbox{$z_\lm=-\X$}. This is clearly shown in figure~\ref{figA1}, where the universal shape of the closest approach curve is compared to a universal shape \mbox{$\bar{z}=\bar{\rho}^2/8-2$} of the scattering shadow from (\ref{shadfix}).

In addition to this demonstration, one could always insist on identifying some trajectory that under an angle $\tilde{\theta}(\R_0)$ comes closer to a target than the trajectory defined by $\R_0$, whose own distance to a target is minimized under the same angle\footnote{
 One such trajectory is certainly be the one that comes closest to a target, whose own impact parameter is $\tilde{\R}_0[\tilde{\theta}(\R_0)]$. Indeed, it is straightforward to show that for a given $\R_0$ the trajectory with $\tilde{\R}_0$ always comes closer: \mbox{$r\{\tilde{\theta}(\R_0);\tilde{\R}_0[\tilde{\theta}(\R_0)]\} \le r[\tilde{\theta}(\R_0);\R_0]$} or, equivalently: \mbox{$r[\tilde{\theta};\tilde{\R}_0(\tilde{\theta})] \le r[\tilde{\theta};\R_0(\tilde{\theta})]$}. Using~(\ref{rmin}) and (\ref{rmin_th})---and keeping in mind in manipulating the inequality that for \mbox{$\tilde{\theta}>\pi/2$} the term $\cos\tilde{\theta}$ is negative---this boils down to showing:
\begin{equation*}
\frac{2\X}{\sin^2(\tilde{\theta}/2)}\le\X\frac{\cos\tilde{\theta}-1}{\cos\tilde{\theta}} \quad\Leftrightarrow\quad (1+\cos\tilde{\theta})^2\ge0.
\end{equation*}
Since the left inequality is equivalent to the right one, and the right one is true for any $\tilde{\theta}$, so is true the initial claim.
}. The problem boils down to finding an impact parameter $\R$ for which the distance from (\ref{masterap}) is smaller than for $\R_0$:
\begin{equation}
r[\tilde{\theta}(\R_0),\R]<r[\tilde{\theta}(\R_0),\R_0].
\label{ineq}
\end{equation}
With the help of trigonometric identities\footnote{
Trigonometric identities in question ultimately yield: 
\begin{equation*}
\sin\left[\frac{\pi}{2}+\atan\frac{\X}{\R_0}-\atan\frac{\X}{\R}\right]=\frac{\X^2+\R_0\R}{\sqrt{(\X^2+\R_0^2)(\X^2+\R^2)}}.
\end{equation*}
} (\ref{ineq}) reduces to a quadratic inequality:
\begin{equation}
\R^2\sqrt{\X^2+\R_0^2}-\R\frac{\R_0^3}{\sqrt{\X^2+\R_0^2}-\X}+\X\R_0^2<0
\end{equation}
to be solved for $\R$. There are two boundaries to this inequality:
\begin{align}
&\R_+=\R_0,\\
&\R_-=\frac{\X\R_0}{\sqrt{\X^2+\R_0^2}},
\end{align}
meaning that, under an angle $\tilde{\theta}(\R_0)$, any trajectory with an impact parameter \mbox{$\R$ such that $\R_-<\R<\R_+$}, comes closer to the target than the trajectory with the impact parameter $\R_0$. The trajectory yielding the absolute minimum distance among all trajectories---the one with an impact parameter $\tilde{\R}_0[\tilde{\theta}(\R_0)]$---is certainly to be found within this range. In fact, plugging (\ref{tilde_th}) into (\ref{tilde}), with the help of a very useful identity \mbox{$\tan(\pi/4+x/2)=\tan x+1/\cos x$}, it can be shown that its impact parameter is precisely the harmonic mean between the two boundaries:
\begin{equation}
\frac{1}{\tilde{\R}_0[\tilde{\theta}(\R_0)]}=\frac{1}{2}\left(\frac{1}{\R_+}+\frac{1}{\R_-}\right).
\end{equation}
Since $\R_+=\R_0$, any trajectory satisfying (\ref{ineq}) has a smaller impact parameter than $\R_0$, hence: \mbox{$\tilde{\R}_0[\tilde{\theta}(\R_0)]\le\R_0$}.

\begin{figure}[t!]
\centering
\includegraphics[width=0.5\textwidth,keepaspectratio]{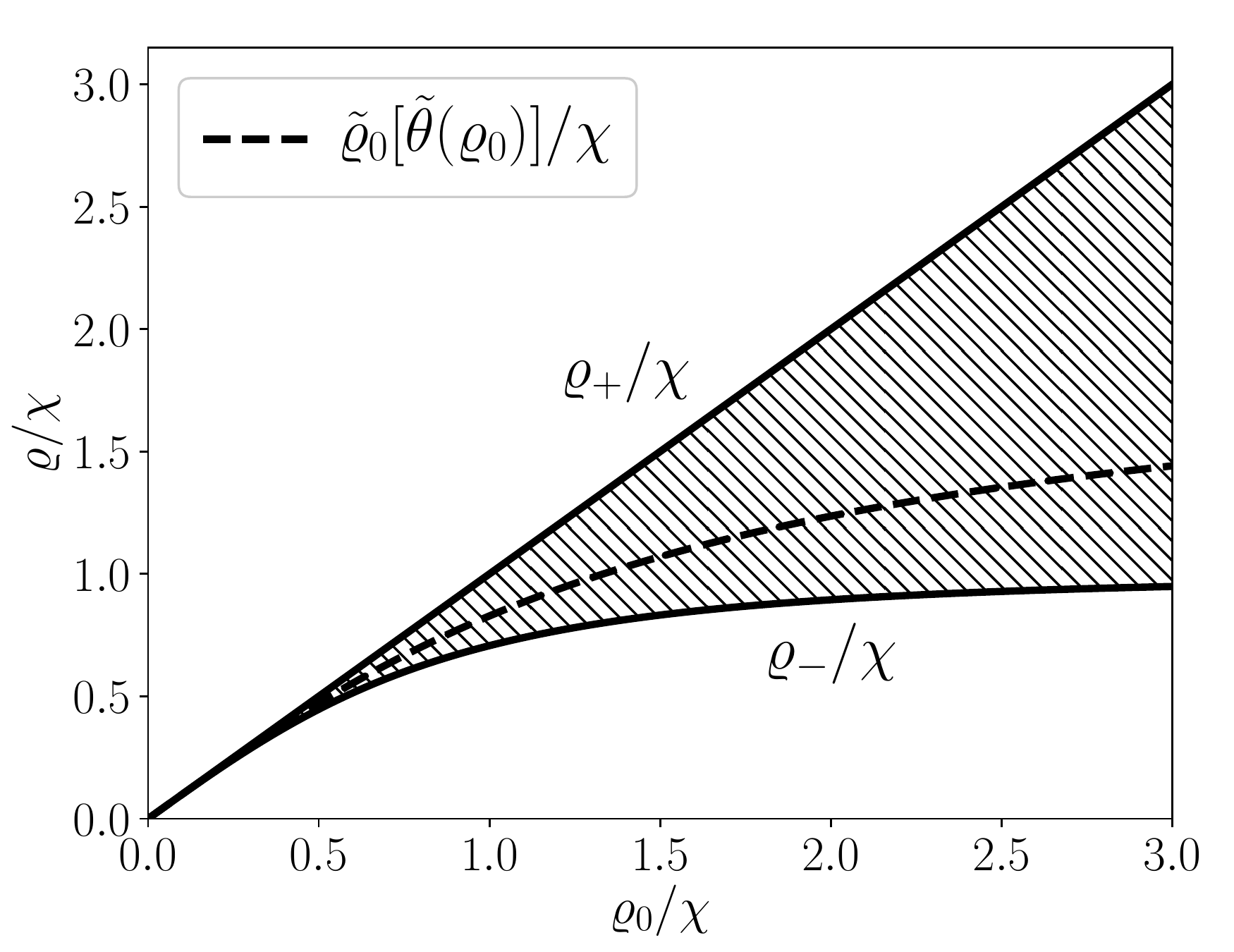}
\pull
\caption{Range of impact parameters $\R$ (shaded area) for which, under an angle $\tilde{\theta}(\R_0)$, the projectile trajectory in a fixed-target frame comes closer to the target than the trajectory with $\R_0$, whose own distance to the target is minimized under the same angle. The impact parameters \mbox{$\tilde{\R}_0[\tilde{\theta}(\R_0)]$} that minimize the projectile-target distance are also shown (dashed line), being a harmonic mean between $\varrho_+$ and $\varrho_-$.\pullc}
\label{figA2}
\end{figure}

Figure~\ref{figA2} shows an exact range of impact parameters from which one may select sought trajectories. By now it should not be at all surprising that these boundaries also take the universal form in scaled coordinates \mbox{$\bar{\R}_{\pm,0}=\R_{\pm,0}/\X$} so that: \mbox{$\bar{\R}_+=\bar{\R}_0$} and \mbox{$\bar{\R}_-=\bar{\R}_0/\sqrt{\bar{\R}_0^2+1}$}. One may also note that for large values of $\R_0$ the nontrivial boundary $\R_-$ saturates: \mbox{$\lim_{\R_0\to\infty}\R_-=\X$}. Therefore, for the reference trajectories sufficiently distant from the target---which are barely deflected and whose point of the closest approach is roughly under \mbox{$\tilde{\theta}(\R_0)\approx\pi/2$}---almost all trajectories with a smaller impact parameter ($\R<\R_0$) come closer to a target, except for those with \mbox{$\R\lessapprox\X$}, that are deflected backwards even before reaching an angle \mbox{$\theta\approx\pi/2$}.

\begin{figure}[t!]
\centering
\includegraphics[width=0.5\textwidth,keepaspectratio]{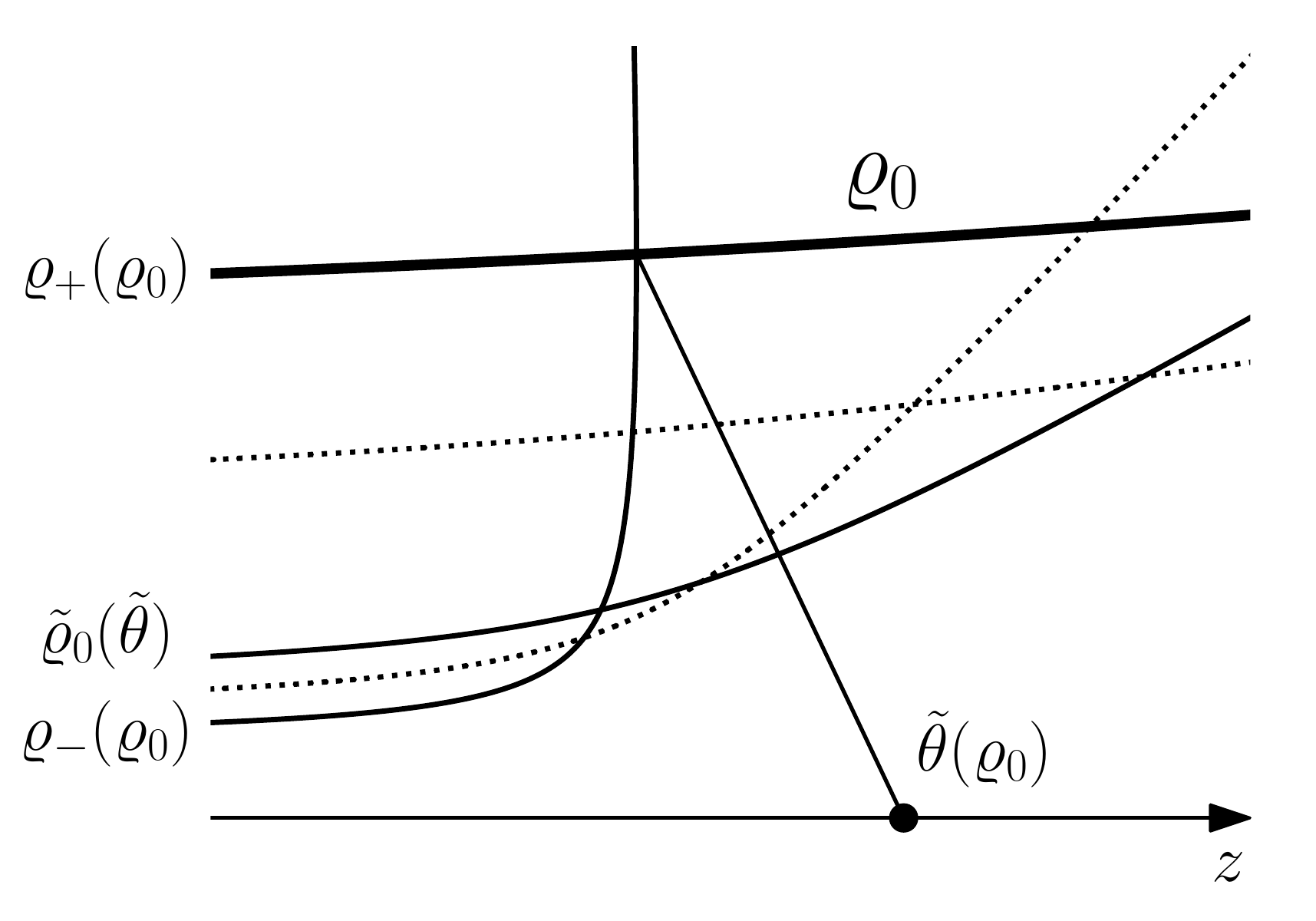}
\pull
\caption{Several trajectories related to the reference one (with an impact parameter $\R_0$), which under an angle $\tilde{\theta}(\R_0)$ reaches the point of the closest approach. See the main text for a full description. The vertical and horizontal scales are not equal, causing the connecting line, depicting $\tilde{\theta}(\R_0)$, not to appear orthogonal to the reference trajectory at an intersection point.\pullc}
\label{figA3}
\end{figure}

Figure~\ref{figA3} shows the relation between various trajectories. The reference trajectory is the one with the impact parameter $\R_0$. Alongside the trajectory with $\R_+(\R_0)$, which is identical to $\R_0$, the one with $\R_-(\R_0)$ features the same target-projectile distance under an angle $\tilde{\theta}(\R_0)$ as the reference trajectory. The trajectory minimizing the target-projectile distance, under the same angle, is also shown: the one with $\tilde{\R}_0(\tilde{\theta})$. Dashed trajectories are examples of those that come closer to the target than the reference trajectory, yet do not minimize that distance. One of them is purposely taken from the range \mbox{$\langle\R_-,\tilde{\R}_0\rangle$}, the other from \mbox{$\langle\tilde{\R}_0,\R_+\rangle$}, i.e. each one from a different side of the distance-minimizing one. It should be noted that for purposes of managing the figure dimensions, the horizontal and vertical scale are not equal. Consequently, the connecting line (depicting an angle $\tilde{\theta}$) does not appear to be orthogonal to the reference trajectory ($\R_0$) at the intersection point (the point of the closest approach), as it should appear if the scales were the same.

\section{Intersection of shadows}

\begin{figure}[t!]
\centering
\includegraphics[width=0.5\textwidth,keepaspectratio]{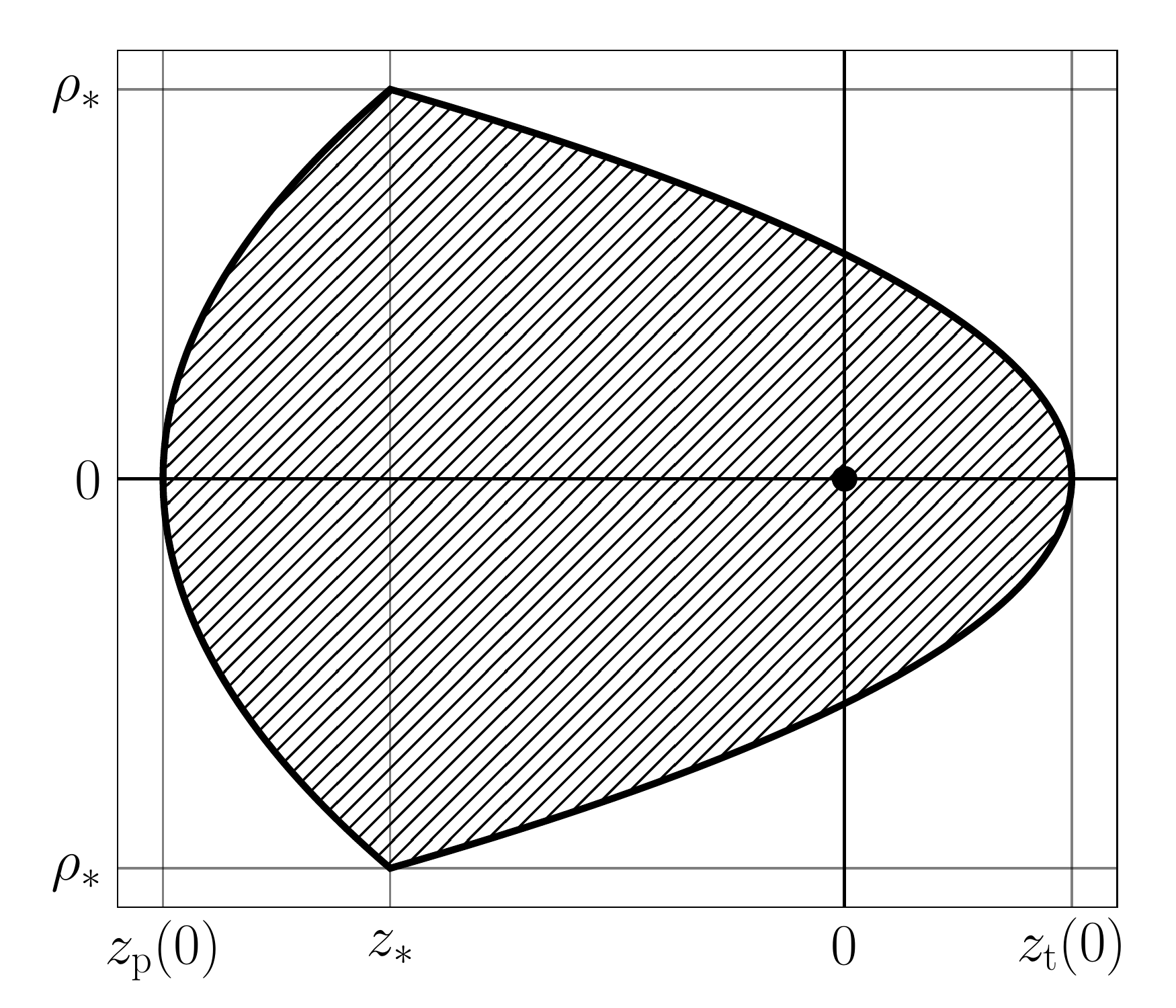}
\pull
\caption{Intersection of the projectile (from left) and the target (from right) shadows in a center-of-mass frame, for $\eta_\pro=0.25$, i.e. $\eta_\tar=0.75$. The portion of space corresponding to a shaded area is entirely shielded from any particle trajectory. The focal points of both shadows coincide and are situated at the origin of the coordinate frame (the dot).\pullc}
\label{figB}
\end{figure}

As seen from figure~\ref{fig1}, in a fixed-target frame an unlimited portion of space is shielded from admitting either the projectile or target trajectories. In a center-of-mass frame each of the two shadows also shields an unlimited portion of space, but only from a corresponding particle. As opposed to that, only a limited portion of space is shielded from \textit{both} the projectile \textit{and} the target. This vacuous portion of space is limited by the intersection of their respective shadows and provides an opportunity for some instructive geometric calculations. Figure~\ref{figB} shows an example of this intersection for \mbox{$\eta_\pro=0.25$} (i.e. \mbox{$\eta_\tar=0.75$} or, equivalently, \mbox{$m_\tar=3m_\pro$}).

In order to calculate the geometric parameters of this intersection, we first need to identify several relevant reference points. The first two are the positions of the shadow vertices along the $z$-axis. From~(\ref{compro}) and (\ref{comtar}) they are trivially read out as:
\begin{equation}
z_\pro(0)=-2\eta_\tar\X \quad\text{and}\quad z_\tar(0)=2\eta_\pro\X.
\end{equation}
The remaining values are the paraboloids' intersection coordinates $\rho_\ast$ and $z_\ast$, which are easily determined by solving the equation \mbox{$z_\pro(\rho_\ast)=z_\tar(\rho_\ast)$}, thus obtaining:
\begin{equation}
\rho_\ast=4\X\sqrt{\eta_\pro\eta_\tar} \quad\text{and}\quad z_\ast=2\X(\eta_\pro-\eta_\tar).
\end{equation}

\begin{comment}
\begin{align}
&z_\pro(0)=-2\eta_\tar\X,\\
&z_\tar(0)=2\eta_\pro\X.
\end{align}

\begin{align}
&\rho_\ast=4\X\sqrt{\eta_\pro\eta_\tar},\\
&z_\ast=2\X(\eta_\pro-\eta_\tar).
\end{align}
\end{comment}

Let us first consider a Rutherford scattering in a two-dimensional space. This space is equivalent to any plane containing the $z$-axis of a full three-dimensional space, and is exactly represented by an example from figure~\ref{figB}. We are interested in calculating the perimeter $P$ and the area $A$ of this two-dimensional intersection of \textit{parabolic} shadows. Returning to a full three-dimensional space, we are interested in calculating two additional, analogous values: the surface area $S$ and he volume $V$ of the three-dimensional intersection of \textit{paraboloidal} shadows.

Starting with the perimeter $P$, we first note that an infinitesimal length element $\D l_{\pro,\tar}$ along any of the two parabolas is given by \mbox{$(\D l_{\pro,\tar})^2=(\D\rho_{\pro,\tar})^2+(\D z_{\pro,\tar})^2$}, so that depending on the choice of an independent integration variable, we may either write \mbox{$\D l_{\pro,\tar}=\D\rho_{\pro,\tar}\sqrt{1+(\D z_{\pro,\tar}/\D\rho_{\pro,\tar})^2}$} or \mbox{$\D l_{\pro,\tar}=\D z_{\pro,\tar}\sqrt{1+(\D\rho_{\pro,\tar}/\D z_{\pro,\tar})^2}$}. In order to calculate the newly appearing derivatives as functions of appropriate coordinates, one may use already available explicit forms \mbox{$z_{\pro,\tar}(\rho_{\pro,\tar})$} from~(\ref{compro}) and (\ref{comtar}). Alternatively, one may first wish to calculate inverse relations \mbox{$\rho_{\pro,\tar}(z_{\pro,\tar})$}---which is done easily enough---and proceed with an integration over $z_{\pro,\tar}$. Therefore, we have these equivalent approaches to calculating $P$:
\begin{align}
\begin{split}
P&=2\sum_{\idx=\pro,\tar}\int_0^{\rho_\ast}\sqrt{1+\left(\frac{\D z_\idx}{\D \rho_\idx}\right)^2} \D\rho_\idx\\
&=2\sum_{\idx=\pro,\tar}\left|\int_{z_\ast}^{z_\idx(0)}\sqrt{1+\left(\frac{\D\rho_\idx}{\D z_\idx}\right)^2} \D z_\idx\right|,
\end{split}
\end{align}
and it is an instructive exercise to show that both procedures indeed yield (as they must!) the same result. In calculating the area $A$ closed by two parabolas, one may again select an independent integration variable, which is equivalent to selecting a particular order of integration when a full double integral is written down. Immediately solving the first, trivial of the two integrations, we may again select one of the two finishing procedures:
\begin{align}
\begin{split}
A&=2\sum_{\idx=\pro,\tar}\left|\int_0^{\rho_\ast}[z_\idx(\rho_\idx)-z_\ast]\D\rho_\idx\right|\\
&=2\sum_{\idx=\pro,\tar}\left|\int_{z_\ast}^{z_\idx(0)} \rho_\idx(z_\idx) \D z_\idx\right|.
\end{split}
\end{align}
These integrals may very easily be set in place just by observing the geometry from figure~\ref{figB}. Similarly, for a surface area $S$ of a three-dimensional intersection, the basis for an integration is $2\pi\rho_{\pro,\tar}\D l_{\pro,\tar}$ (where the first of the two integrals goes over the azimuthal angle~$\varphi_{\pro,\tar}$, yielding $2\pi$), so that:
\begin{align}
\begin{split}
S&=2\pi\sum_{\idx=\pro,\tar}\int_{0}^{\rho_\ast}\rho_\idx\sqrt{1+\left(\frac{\D z_\idx}{\D \rho_\idx}\right)^2} \D \rho_\idx\\
&=2\pi\sum_{\idx=\pro,\tar}\left|\int_{z_\ast}^{z_\idx(0)}\rho_\idx(z_\idx)\sqrt{1+\left(\frac{\D\rho_\idx}{\D z_\idx}\right)^2} \D z_\idx\right|.
\end{split}
\end{align}
Finally, for he volume $V$ we have:
\begin{align}
\begin{split}
V&=2\pi\sum_{\idx=\pro,\tar}\left|\int_{0}^{\rho_\ast} \rho_\idx [z_\idx(\rho_\idx)-z_\ast] \D \rho_\idx\right|\\
&=\pi\sum_{\idx=\pro,\tar}\left|\int_{z_\ast}^{z_\idx(0)} \rho_\idx^2(z_\idx) \D z_\idx\right|.
\end{split}
\end{align}
Though it is instructive to carry out both types of integrations and for both particles separately, in reality it is entirely sufficient not only to perform just one type of integration (either over $\rho_{\pro,\tar}$ or $z_{\pro,\tar}$), but just for one of the particles (either for the projectile or the target). The reason is the symmetry between the projectile and the target in the center-of-mass frame, as they roles can always be interchanged by a simple interchange of indices $\pro\leftrightarrow\tar$ throughout all the expressions. Therefore, if one calculates any of the integrals for one particle, the integral for the other one immediately follows by consistently switching all the indices: $\pro\leftrightarrow\tar$. As a consequence, the final results---as the sums of these two contributions---must be fully symmetric in respect to both particles and invariant under the same index interchange. Indeed, whichever course of calculation is selected, one invariably arrives at:
\begin{align}
&P=4\X\left(\sqrt{\eta_\pro}+\sqrt{\eta_\tar} + \eta_\pro \,\ash\sqrt{\frac{\eta_\tar}{\eta_\pro}}+ \eta_\tar \,\ash\sqrt{\frac{\eta_\pro}{\eta_\tar}}\right),
\label{C}\\
&A=\frac{32}{3}\X^2\sqrt{\eta_\pro\eta_\tar},
\label{A}\\
&S=\frac{32}{3}\X^2\pi \left(\sqrt{\eta_\pro}+\sqrt{\eta_\tar} -\eta_\pro^2-\eta_\tar^2\right),
\label{S}\\
&V=16\X^3\pi\eta_\pro\eta_\tar.
\label{V}
\end{align}

As a closing note, it is worth examining the limiting case of an infinitely heavy target, as in that case the center-of-mass frame coincides with the fixed-target frame. Thus the target stays at rest, \textit{its shadow spanning an entire geometric space}, in a sense of not admitting any target trajectory. We already know from an open projectile shadow (figure~\ref{fig1}) that an infinite portion of space is then shielded from admitting any particle. Therefore, it is a useful exercise to confirm whether or not~(\ref{C})--(\ref{V}) reproduce that limit, as the very limiting case is, strictly speaking, outside their domain\footnote{
The reason for the limiting case of the finite-mass target (\mbox{$m_\tar\to\infty$}) not corresponding to the `exact' case of the infinite mass target (\mbox{$m_\tar=\infty$}) is the following: the initial offset of the finite-mass target from the center-of-mass frame \mbox{$z_\tar^{(\cm)}(m_\tar<\infty;t=0)=\infty$} cannot reproduce, in the limit \mbox{$m_\tar\to\infty$}, the exact offset (at any moment) of the infinite-mass target \mbox{$z_\tar^{(\cm)}(m_\tar=\infty;t)=0$}.
}. In that, it is not sufficient just to set the mass ratios to \mbox{$\eta_\tar=1$} and \mbox{$\eta_\pro=0$}. Rather, one needs to insist that the target mass be infinite ($m_\tar\to\infty$), as opposed to allowing the vanishing projectile mass instead ($m_\pro\to0$, leading to the same set of mass ratios). This is because the Rutherford scattering is sensitive to absolute masses, which is realized through the appearance of the reduced mass $\mu$ within the definition of $\X$ from (\ref{xap}), i.e. from the fact that the reduced mass can not be expressed solely as a function of the mass ratios, but depends on the particular masses: \mbox{$\mu=\eta_\pro \eta_\tar(m_\pro+m_\tar)$}. In examining the limits we are again faced with an issue from (\ref{chieta}), consisting in a caution against naively manipulating just $\eta_\pro$ and $\eta_\tar$ as they explicitly appear in~(\ref{C})--(\ref{V}). Taking into account also their presence within $\X$, the relevant portions of the expressions to be examined take the form:
\begin{align}
&P\propto \notag\\
&\quad\frac{1}{\sqrt{m_\tar  \mu}}+\frac{1}{\sqrt{m_\pro  \mu}}+\frac{1}{m_\tar}\ash\sqrt{\frac{m_\tar}{m_\pro}}+\frac{1}{m_\pro}\ash\sqrt{\frac{m_\pro}{m_\tar}},
\label{Cmass}\\
&A\propto \frac{1}{\mu\sqrt{m_\tar m_\pro}},
\label{Amass}\\
&S\propto \frac{1}{\sqrt{m_\tar \mu^3}}+\frac{1}{\sqrt{m_\pro \mu^3}}-\frac{1}{m_\tar^2}-\frac{1}{m_\pro^2},
\label{Smass}\\
&V\propto \frac{1}{m_\tar m_\pro  \mu}.
\label{Vmass}
\end{align}
Since \mbox{$\lim_{m_\tar\to\infty}(\ash\sqrt{m_\tar/m_\pro})/m_\tar=0$} (which is easily shown by the L'H\^{o}pital's rule) and \mbox{$\lim_{m_\tar\to\infty}\mu=m_\pro$}, we see that in the limit $m_\tar\to\infty$ most of the terms vanish, except for the second term from (\ref{Cmass}), together with the second and the fourth term from (\ref{Smass}). As their placement corresponds to that of the terms from~(\ref{C}) and (\ref{S})---with \mbox{$\eta_\tar=1$} in the same limit---we may immediately write:
\begin{align}
&\lim_{m_\tar\to\infty} P=4\X,\\
&\lim_{m_\tar\to\infty} A=\lim_{m_\tar\to\infty} S=\lim_{m_\tar\to\infty} V=0.
\end{align}
These results are easy to understand, as in the limiting case of (\ref{comtar}) the target shadow in the center-of-mass frame becomes an infinitely narrow paraboloid, intersecting the projectile shadow before it even detaches from the $z$-axis. Thus the intersection of such shadows forms a geometric shape akin to a one-dimensional line segment, of vanishing volume and any areal property, but still of finite one-dimensional features such as the perimeter. As we had set to investigate, we see now that this limiting case does not coincide with the `exact' case of the infinitely massive target, since the qualitative change in dynamics takes place---the target stays at rest, exempting an entire geometric space from its trajectories. Therefore, the true value for each of the calculated geometric properties \mbox{$\Gamma\in\{P,A,S,V\}$} diverges, in a sense \mbox{$\Gamma(m_\tar=\infty)=\infty$}, so that we have an incongruity: \mbox{$\Gamma(m_\tar\to\infty)\neq\Gamma(m_\tar=\infty)$}.

\clearpage

\end{document}